\documentclass[%
superscriptaddress,
reprint,
preprintnumbers,
 amsmath,amssymb,
aps,
]{revtex4-2}

\usepackage{graphicx}
\usepackage{dcolumn}
\usepackage{bm}
\usepackage{hyperref}
\usepackage[mathlines]{lineno}

\usepackage{amsmath}
\usepackage{amsfonts}
\usepackage{amssymb}
\usepackage{graphicx}
\usepackage{mathrsfs}
\usepackage{times}
\usepackage{color}
\usepackage{psfrag}
\usepackage{natbib}
\definecolor{violet}{rgb}{0.5,0,0.5}
\definecolor{blue2}{rgb}{0.5,0,1}
\usepackage{setspace}
\usepackage{graphicx,subfig}
\usepackage{array,calc}
\usepackage{braket}

\newcommand{\di}{i}

\usepackage[justification=raggedright]{caption}

\begin{document}


\title{Wave beaming and diffraction in quasicrystalline elastic metamaterial plates}

\author{Danilo Beli} \email{dbeli@usp.br, beli.danilo@gmail.com}
\affiliation{Department of Aeronautical Engineering, Sao Carlos School of Engineering, University of Sao Paulo, Brazil}
\author{Matheus Inguaggiato Nora Rosa} \email{matheus.rosa@colorado.edu}
\affiliation{Department of Mechanical Engineering, College of Engineering and Applied Science, University of Colorado Boulder, USA} 
\author{Carlos De Marqui Jr.} \email{demarqui@sc.usp.br}
\affiliation{Department of Aeronautical Engineering, Sao Carlos School of Engineering, University of Sao Paulo, Brazil}
\author{Massimo Ruzzene} \email{massimo.ruzzene@colorado.edu}
\affiliation{Department of Mechanical Engineering, College of Engineering and Applied Science, University of Colorado Boulder, USA}

\date{\today}

\begin{abstract}
In this paper, we present numerical and experimental evidence of  directional wave behavior, i.e. beaming and diffraction, along high-order rotational symmetries of quasicrystalline elastic metamaterial plates. These structures are obtained by growing pillars on an elastic plate following a particular rotational symmetry arrangement, such as 8-fold and 10-fold rotational symmetries, as enforced by a design procedure in reciprocal space. We estimate the dispersion properties of the waves propagating in the plates through Fourier transformation of transient wave-fields. The procedure identifies, both numerically and experimentally, the existence of anisotropic bands characterized by high energy density at isolated regions in reciprocal space that follow their higher order rotational symmetry. Specific directional behavior is showcased at the identified frequency bands, such as wave beaming and diffraction. This work expands the wave directionality phenomena beyond the symmetries of periodic configurations (e.g., 4-fold and 6-fold), and opens new possibilities for applications involving the unusual high-order wave features of the quasicrystals such as superior guiding, focusing, sensing and imaging.
\end{abstract}

\maketitle



\section{Introduction}

Periodic configurations have dominated the designs of phononic crystals and metamaterials in the past decades. Although the wave features of architected materials are usually associated with their crystalline symmetry and translational periodicity \cite{Joannopoulos1997, Hussein2014, Christensen2015}, non-periodic configurations have also been explored in order to achieve various wave manipulation capabilities \cite{Ma2016, Cummer2016, Assouar2018}. For example, trivial defects and topological interfaces have been employed for flexible wave guiding \cite{Yang2015TopAcoustics, Bandres2016, Fremling2020}, and also, spatially correlated unit cells (e.g., rainbow) or disorder have been shown the ability to trap waves and broadband vibration attenuation \cite{Celli2019, Beli2019}. In this context, quasiperiodic phononic configurations or quasicrystals have emerged as relevant candidates for unusual wave phenomena \cite{Kraus2016}. Their configurations in physical space lack translational periodicity, but long-range order as well as high-order rotational symmetries are present \cite{Vardeny2013}. The unique symmetries of quasicrystals are revealed by their exotic sharp Bragg diffraction patterns, first experimentally observed by Shechtman et al. \cite{Shechtman1984} and theoretically reported in the pioneering work of Levine and Steinhardt \cite{LevineSteinhardt1984}. Investigation of properties arising from such unique symmetries has resulted in novel applications such as lasing \cite{Notomi2004, Luo2012}, superior sensing and imaging \cite{Han2020}, guiding and bending of waves \cite{An2020}, super-focusing \cite{Vardeny2013}, superconductivity \cite{Sakai2019} and topological wave transport \cite{apigo2019observation, ni2019observation}. 

In the context of elastic materials, the higher order rotational symmetries of quasicrystals has been shown to induce nearly isotropic stiffness properties in lattice  structures~\cite{Wang2020,Chen2020} and continuum elastic composites \cite{Beli2021} which results in omnidirectional wave propagation at low frequencies. Their mechanical isotropy can also make them less sensitive to geometric and material variabilities \cite{Chen2021}. Similar to the phononic crystals, band gaps or pseudo gaps have also been observed in phononic quasicrystals \cite{Florescu2009}; their bands, however, usually split in several mini bands due to their fractal nature \cite{Kaliteevski2000}. In addition, the dispersion properties of quasicrystals are associated with their representation in wave number space, which has motivated a pseudo Brillouin zone definition \cite{Luo2012, Rotenberg2000}. The approximated dispersion, however, has been computed only for simple quasiperiodic lattices \cite{Gambaudo2014, Fuchs2018}, and, in most cases, assuming a periodic approximation. Other works have investigated the waveguiding capabilities of quasicrystals with~\cite{Cheng1999} or without~\cite{Wang2006, Sinelnik2020} defects. More recently, quasiperiodic arrangements have also been employed to pursue higher dimensional topological features, which emanate from the existence of additional parameters (such as the phason) and provide opportunities for topological states with corner localization, lower dimension guiding and pumping \cite{Fuchs2016Top,apigo2018topological, Rosa2019, zhou2019topological, Chen2020Top,cheng2020experimental, Spurrier2020,xia2020topological,gupta2020dynamics,
xia2021experimental,rosa2021exploring,rosa2021topological,Koshino2022}.

\begin{figure*} 
	\centering
	{\includegraphics[scale = 0.510]{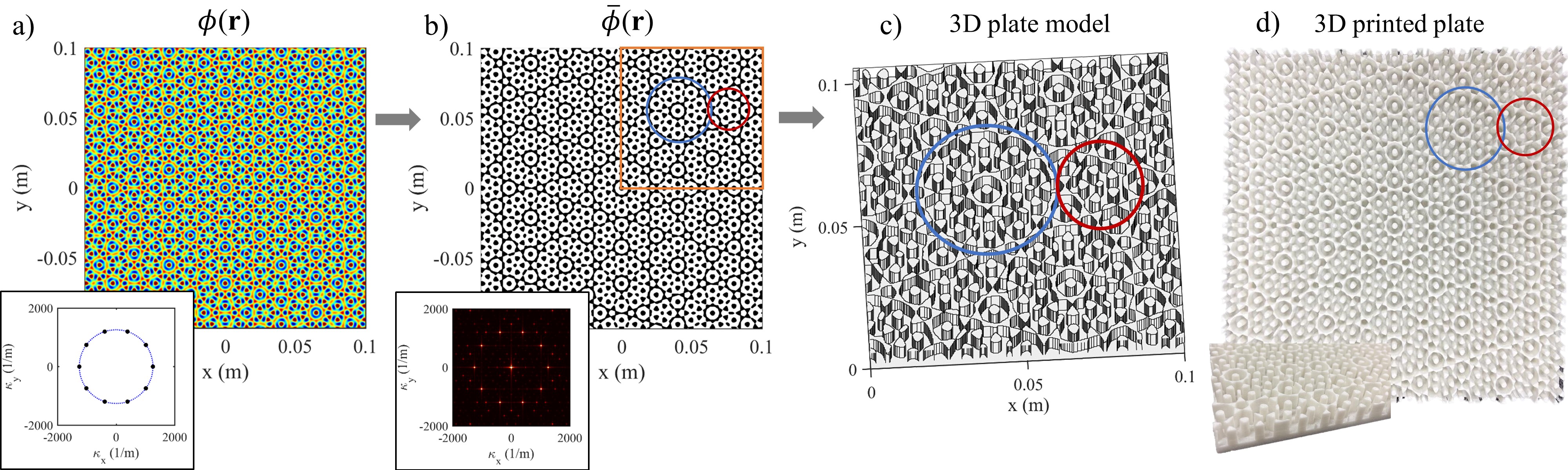}} 
	\caption{\label{fig:figure1} Design strategy for the 10-fold symmetry with ${\tt vf} = 0.30$: two-dimensional physical distribution by assigning 10 Bragg peaks (a), two-phase distribution after applying the threshold procedure and its Fourier transform (i.e., diffraction pattern) (b): phase $\tt A$ (white) and phase $\tt B$ (black). Three-dimensional quasiperiodic plate obtained by extruding phase $\tt B$ towards z direction, which produces the geometry in (c) and the correspondent 3D printed quasiperiodic plate (d).} 
\end{figure*} 

Despite the recent interest on the dynamics of quasiperiodic systems, their dispersion and wave directionality properties are still largely unexplored or scarce. Indeed, wave directionality finds applications that involve wave filtering and guiding, ultrasonic therapy and imaging, antennas, sensors and lenses; however, the wave fronts are currently restricted to known crystallographic symmetries (e.g., bilayer, square and hexagonal) \cite{Ruzzene2003, RuzzeneScarpa2005, Phani2006, Gonella2008, Trainiti2016, Beli2018, Foehr2018, Rosi2019, Grabec2020}. As we illustrate herein, quasiperiodic configurations expand the wave directionality and beaming possibilities beyond the symmetries of periodic arrangements. Inspired by previous work on quasicrystalline composites~\cite{Beli2021}, we investigate the spectral properties and directional wave behavior of elastic plates whose higher order rotational symmetries are enforced through a design procedure in reciprocal space. The directional behavior along higher-order rotationally symmetric directions (i.e. 8-fold and 10-fold) is illustrated in simulations and confirmed by experiments conducted on samples fabricated through additive manufacturing. The results of this paper highlight unexplored features of the wave behavior in quasicrystalline media, i.e. wave directionality in high-order rotational symmetries, and open new possibilities for practical implementations in structural components and wave devices. 


\section{Quasicrystalline plates design}

The strategy to design the quasiperiodic elastic metamaterial plates is based on the geometric representation in 2D wave number space \cite{Lubensky1988, Widom2008, Beli2021}. A continuum distribution in physical space $\phi({\bf r})$, with ${\bf r} = [x,y] \in \mathcal{R}^2$, is defined by directly assigning $N$ Bragg peaks in reciprocal space (${\bf k} = [k_x,k_y] \in \mathcal{R}^2$) as points in the two-dimensional Fourier spectra \cite{Vardeny2013}. These Bragg peaks are angularly spaced by $\theta_N = 2\pi/N$ over a circle of fundamental wave number $k_0$. In this work, only even number of peaks is considered to guarantee a real distribution in physical space. Therefore, reciprocal and physical spaces can be expressed, respectively, as:
\begin{equation}
\hat{\phi}({\bf k}) = \sum_{n=0}^{N-1}\delta({\bf k}-{\bf k}_n) \quad \text{and} \quad
\phi({\bf r}) = \sum_{n = 0}^{N-1} e^{\di {\bf k}_n \cdot {\bf r}},
\label{eq:eq1}
\end{equation}
where $\delta$ is the delta function that locates the wave number ${\bf k}_n$ of each Bragg peak; moreover, ${\bf k}_n = k_0 [\cos \left( n \theta_N \right), \,  \sin \left( n \theta_N \right)]$, with $ n = 0,..., N-1$ and $k_0 = 2\pi/\lambda_0$ is the radius of the design circle in reciprocal space, where $\lambda_0$ is the fundamental wavelength. In this design strategy, a single parameter $N$ defines the rotational symmetry of the distribution in physical space, which leads to periodic distributions (1D bilayer for $N=2$, square pattern for $N=4$ and hexagonal pattern for $N=6$) or quasiperiodic distributions with rotationally $N$-fold symmetry such as the 8-fold and 10-fold.

For practical implementations in elastic continuum structures, a two-phase distribution is desirable, and hence, a threshold procedure is applied to the real continuum field. This new distribution $\bar{\phi}({\bf r})$ assumes only two phases, which are produced by comparing the local field level to a chosen level $\bar{\phi}_{0}$: a phase $\tt A$ is defined for $\phi({\bf r}) \leq \bar{\phi}_{0}$ and a phase $\tt B$ is defined for $\phi({\bf r}) > \bar{\phi}_{0}$. Based on the phase ratio, a volume (or filling) fraction is defined by ${\tt vf} = {v_{\tt B}/(v_{\tt A}+v_{\tt B})}$. Herein, the quasicrystalline metamaterial plate is designed using a single material with geometric thickness modulation given by $\bar{\phi} \left( {\bf r} \right)$, where a flat plate (phase $\tt A$) is partially covered in one side by pillars (phase $\tt B$). These geometries can be conveniently manufactured using regular additive manufacturing technologies. Moreover, experimental observations can be performed through vibration measurements on the flat side. Figure \ref{fig:figure1} summarizes the design process, from the choice of Bragg peaks in the reciprocal space to the three-dimensional printed plate with 10-fold symmetry and ${\tt vf}=0.30$ (check the Supplemental Materials (SM) for the 4-, 6-, and 8-fold metamaterial plates). Reference \cite{Beli2021} also details this design strategy for other fold symmetries and volume fractions considering two-dimensional domains with in-plane properties modulation, i.e. steps (a-b) on Fig. \ref{fig:figure1}. 

The plates have a square domain in the $xy$-plane of size $L = 0.2$ m and $\lambda_{0} = 5$ mm, and their modulated thickness in $z$-axis is given by $h({\bf r}) = h_{\tt A} + \bar{\phi}({\bf r}) (h_{\tt B}-h_{\tt A})$, where $h_{\tt A} = 4$ mm and $h_{\tt B} = 12$ mm. In addition, they are manufactured using selective laser sintering (SLS) process and polymer nylon 12, with nominal elastic properties: mass density $\rho_n = 1500$ kg/m$^3$, elastic modulus $E_n = 5$ GPa and Poisson ratio $\nu_n = 0.3$.


\begin{figure} [h!]
	\centering
	{\includegraphics[scale = 0.500]{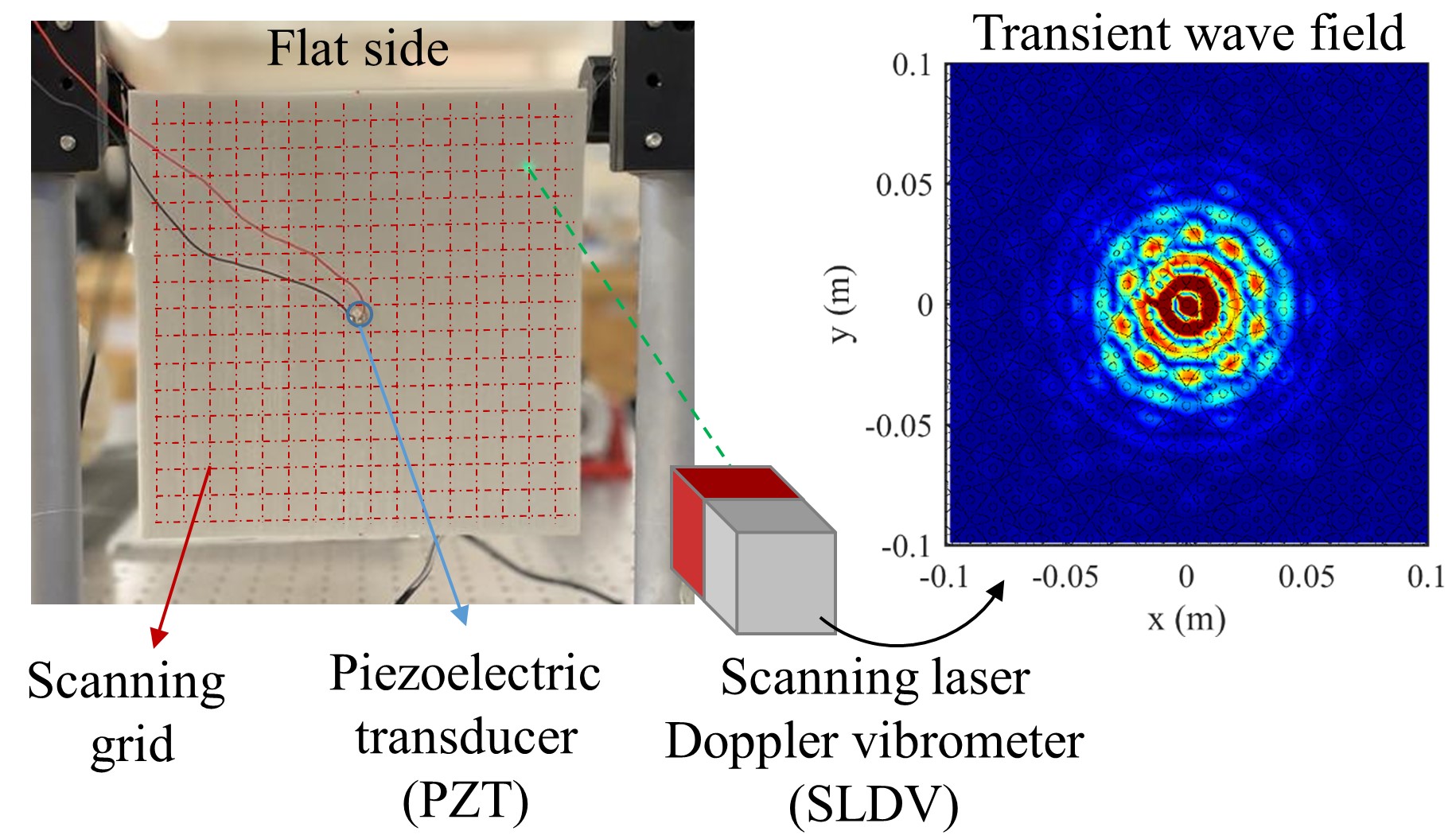}} 
	\caption{\label{fig:figure1b} Experimental set-up for time response observations, the PZT transducer excitation and the SLDV measurements are placed on the flat side of the plate, which recovers the two-dimensional wave field for each instant of time.} 
\end{figure}

\begin{figure*} 
	{\includegraphics[scale = 0.50]{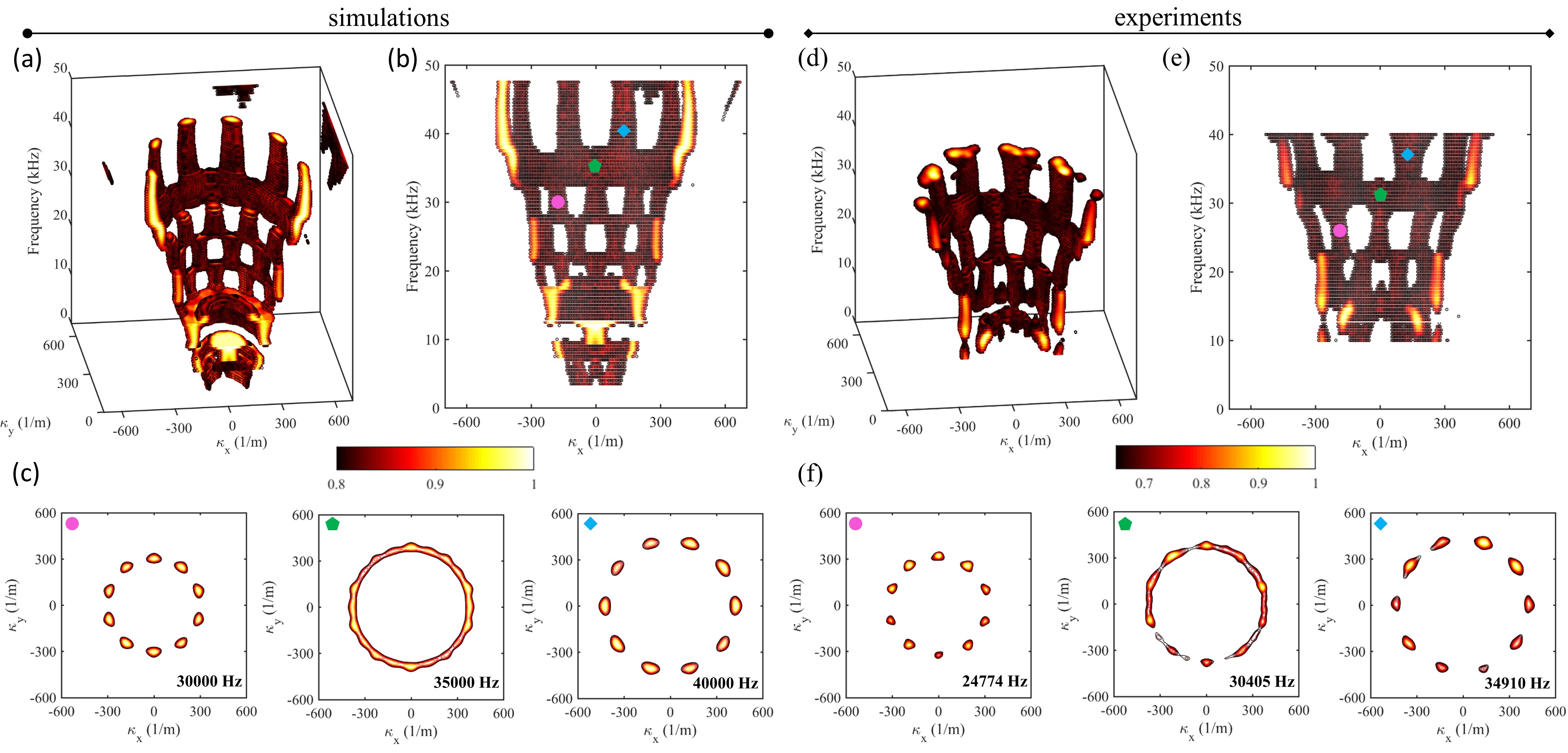}} 
	\caption {Dispersion properties of the 10-fold quasicrystal plate obatined by numerical simulations (a-c) and experiments (d-f). Approximated dispersion surfaces obtained by 3D Fourier transform of the time response: 3D view (a, d) and sectional view on $\kappa_x \omega$ plane (b, e) . Contours at specific frequencies showing the transition between two bands twisted by $\theta_N/2$ (c, f).} 
	\label{fig:figure2}
\end{figure*} 


\section{Numerical and experimental methods}

The numerical simulations are conducted using the finite element (FE) approach within the {\tt COMSOL Multiphysics} \textsuperscript{\textregistered} environment, where 3D elastic solid elements with linear strains are employed. The FE discretization results in equations of the form ${\bf M}\ddot{\bf u}({\bf r},t)+{\bf K}{\bf u}({\bf r},t) = {\bf f}({\bf r},t)$, where ${\bf M}$ is the mass matrix, ${\bf K}$ is the stiffness matrix , ${\bf u}$ is the displacement vector, and ${\bf f}$ is the externally applied load vector. The employed meshes comprise 10 elements per wavelength, i.e. $\Delta a = \lambda_0/10$, and for time response simulations, a time-step of $\Delta t = 1/(20f_e)$ is used to appropriately describe the dynamic behavior, where $f_e$ is the excitation frequency in Hz. 

The experimental set-up is shown in Fig. \ref{fig:figure1b}, where free boundary conditions are emulated by suspending the plate by nylon strings. The input excitation is due to a circular piezoelectric transducer (PZT) placed at the center of the flat side. In this work, only the flexural behavior (i.e., bending waves) is considered, and the out-of-plane velocities of a rectangular grid composed of 63-by-63 points on the flat side of the plate are measured by a scanning laser Doppler vibrometer (SLDV), connected to a data acquisition and signal processing unit. 

\begin{figure*}
	\centering
	{\includegraphics[scale= 0.50]{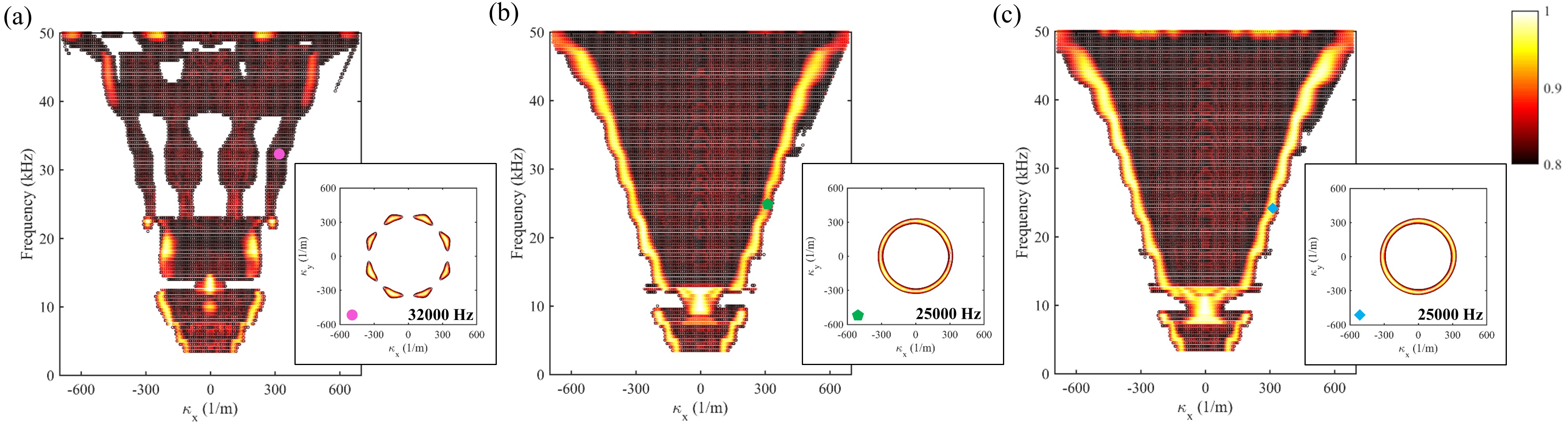}} 
	\caption{\label{fig:figure2b} Numerical approximate dispersion surface with sectional view on $\kappa_x, \omega$ plane for 8-fold (a), 6-fold (b) and 4-fold (c) elastic metamaterial plates with one contour displayed for each case at a specific frequency.} 
\end{figure*}


\section{Dispersion Characterization} 

The dispersion properties are used to understand and predict the dynamic behavior of phononic and metamaterial structures related to wave propagation and manipulation. For periodic materials, the band structure is obtained by enforcing Bloch conditions on a unit cell. However, Bloch-Floquet theory cannot be applied to the present quasicrystalline plates due to their lack of translational periodicity. Instead, we rely on transient wave-fields ${\bf u}(x,y,t)$, and their correspondent 3D-FT (Fourier transform) ${\bf \hat{U}}(k_x,k_y,\omega)$, to estimate the dispersion properties of the quasicrystalline plates. After the general characterization of the estimated dispersion properties presented herein, the following section illustrates specific directional wave behavior associated with the identified anisotropic bands.

In both simulations and experiments, time transient analyses with sinusoidal burst excitation signals are performed with 1-2 cycles for a center frequency of 25 kHz in order to have a broadband excitation to characterize the dispersion in the entire frequency band. Next, 3D-FTs are performed on the displacement fields ${\bf u}(x,y,t)$, providing a representation in reciprocal space where each coordinate $(k_x,k_y,\omega)$ has a wave amplitude ${\bf \hat{U}}$. To reduce effects of unwanted noise at low amplitudes and to improve visualization, the dispersion results are filtered such that only points in the spectrum with higher wave amplitudes are plotted (higher than 0.80 and 0.70 of the maximum value in each frequency for numerical and experimental results, respectively). This approach is validated based on the periodic cases (i.e., 4-fold and 6-fold plates presented in the SM), whose bands and band gaps are readily available from the application of Bloch analysis.

The dispersion behavior of the quasicrystalline plates is first exemplified by the 10-fold case (i.e., $N = 10$), whose design is illustrated in Fig.~\ref{fig:figure1}, and with results summarized in Fig. \ref{fig:figure2}. Overall, a good agreement between numerical simulations and experimental observations has been achieved despite a frequency shift of roughly 5 kHz, which we attribute to uncertainties in the properties of the 3D printed material. The dispersion surfaces for the flexural waves are characterized by a 10-fold rotational symmetry that manifests throughout the majority of the bands, as highlighted by the contours displayed for selected frequencies. In particular, several bands of highly anisotropic contours are identified, which are characterized by 10 separated peaks of high amplitude in reciprocal space forming the 10-fold symmetry. We note that the rotational symmetry twists by $\theta_N/2=18^{o}$ in certain frequency ranges shown in Fig. \ref{fig:figure2}(b), for example around 22, 26 and 35 kHz. One transition example is highlighted by the selected contours of Fig. \ref{fig:figure2}(c), changing from a given 10-fold symmetry arrangement (pink marker), to an almost circular contour (green marker), and then to another 10-fold symmetric arrangement (blue marker), but twisted by $\theta_N/2=18^{o}$ with respect to the previous case (pink marker). This behavior is confirmed by the experimental results of Fig. \ref{fig:figure2}(e,f), and highlights how the dispersion properties of the quasicrystalline plates are characterized by several bands that preserve the $N$-fold symmetry of the design, and may present different anisotropy directions.

A summary of the dispersion results for other symmetry orders are displayed in Fig. \ref{fig:figure2b} and the complete dispersion results are shown in the SM. The behavior of the 8-fold symmetric case is similar to the one observed in Fig. \ref{fig:figure2}, zones of high anisotropy characterized by $N=8$ peaks of high amplitude in reciprocal space are observed. The 4- and 6-fold periodic plates, however, exhibit continuous and highly isotropic bands with almost circular contours, that are due to the identical pillars repeated periodically in space. Therefore, they produce only few and well-defined local-resonant gaps: one around 10 kHz and another around 48 kHz. On the other hand, the quasicrystalline plates are characterized by numerous different pillars, of different resonance frequencies, which are arranged in space according to their higher order rotational symmetry. As a result, the continuous bands of the periodic plates are split into several mini-bands in the quasicrystalline cases, the number of which seems to increase with $N$. Hence, the anisotropic $N$-fold symmetric bands of the quasicrystalline plates seem to emerge from a combination of multiple local resonances that interact and interfere on a higher-order symmetric pattern.


\begin{figure*}
	\centering
	{\includegraphics[scale= 0.520]{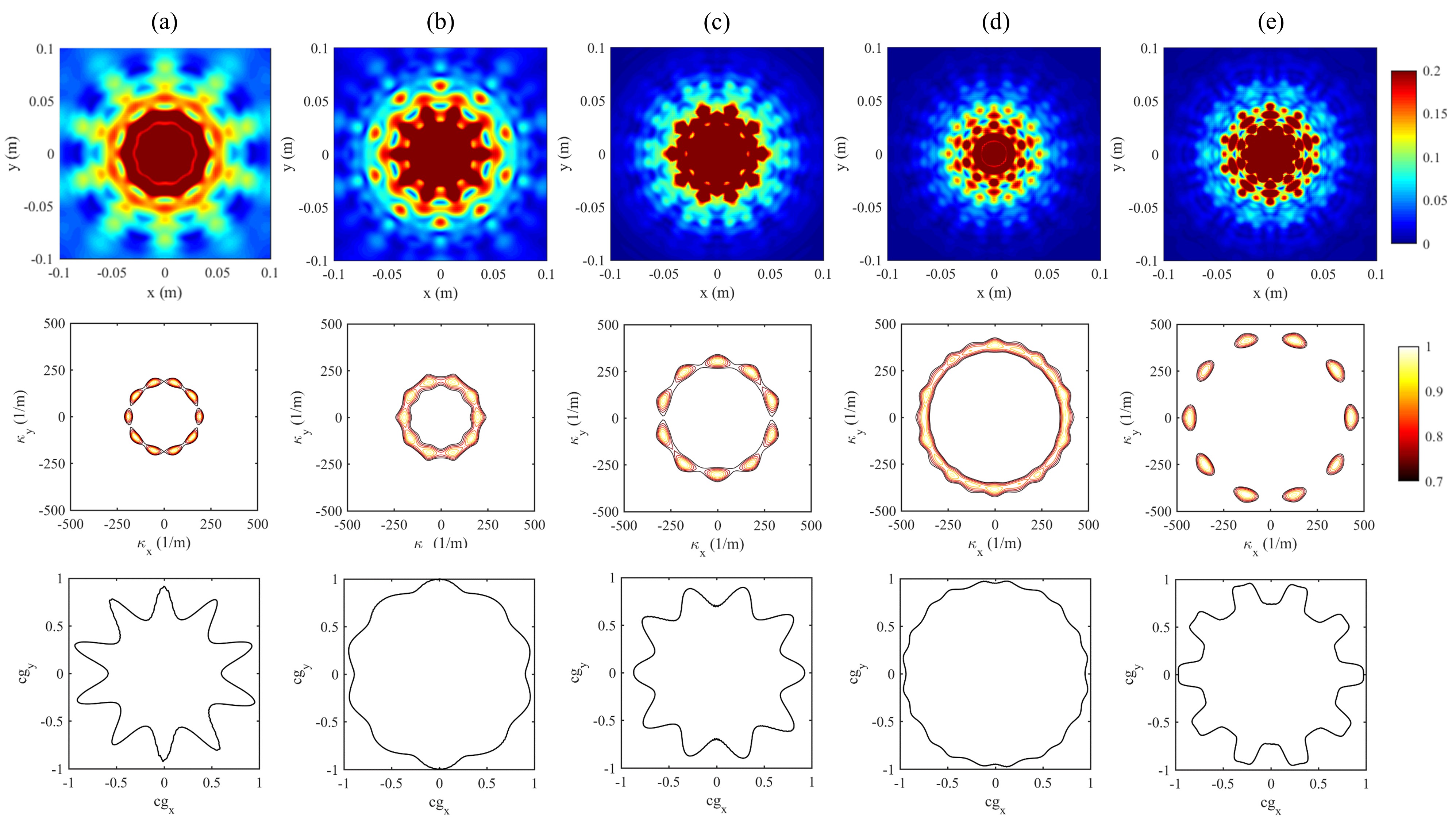}} 
	\caption{\label{fig:figure3b} Directional wave behavior for the 10-fold metamaterial plate at different excitation frequencies: 9.3 kHz (a), 15 kHz (b), 30 kHz (c), 35 kHz (d) and 45 kHz (e). The first row corresponds to the RMS of the wave field averaging across all time, the second row corresponds to the RMS of the wave number contours averaging across all frequencies, and the third row corresponds to the estimated group velocity contour at the center frequency.} 
\end{figure*}

\section{Wave directionality} 

Next, we illustrate in more detail the wave directionality associated with the anisotropic frequency bands identified in the previous section. For such analysis, the transient behavior at selected frequencies is observed by computing the response to narrow-band sinusoidal burst signals (the number of excitation cycles is adjusted for each frequency so that the excitation signal ends as the wave reaches the boundaries of the plate). Figure \ref{fig:figure3b} displays the numerical predictions for the 10-fold metamaterial plate. Each column corresponds to a different excitation frequency in the range from 9.3kHz to 45kHz; the panels on the first row display the root mean square (RMS) of the time response, while the second row displays the RMS wave numbers computed by using the 3D-FT of the response. The directional behavior is further elucidated by estimating the group velocity at the excitation frequencies. For periodic materials, directional wave propagation at a given frequency can be described by the group velocity, i.e. ${\bf c}_g = \nabla_{\kappa} \omega(\kappa)$, where $\omega(\kappa)$ represents the dispersion of the Bloch bands~\cite{Hussein2014}. In the absence of periodicity, we attempt to numerically estimate the group velocities of the quasicrystalline plates based on the dispersion results reported in the previous section. The computation procedure relies on the estimation of the wave number contours for each frequency, their representation on cylindrical coordinates and the derivative computation   through a finite difference. The procedure is fully detailed in the SM, where it is also validated against traditional computations using the Bloch-Floquet theory for periodic configurations. In Fig.~\ref{fig:figure3b}, the third row displays the approximate group velocity contours at the excitation frequency, which are in good agreement with the observed wave-fields. Specifically, at $9.3$kHz (\ref{fig:figure3b}a) the waves propagate preferentially along 10 symmetric directions in a wave-beaming fashion. The Fourier transform shows 10 Fourier peaks that characterize such behavior, while the group velocity further confirms the preferential directions of wave propagation. At the excitation frequency of $15$kHz (\ref{fig:figure3b}b), a transition between two anisotropic bands (\ref{fig:figure3b}a,c), wave propagation is not strongly directional, as confirmed by the almost circular contour in the reciprocal space, and by the smoother group velocity plot. For $30$kHz (\ref{fig:figure3b}c), we observe another directional wave beaming which occurs along directions twisted by $\theta_N/2=18^{o}$ with respect to the case in Fig.\ref{fig:figure3b}(a). Another transition at the excitation frequency of $35$kHz is illustrated in (\ref{fig:figure3b}d), followed by another directional case at $45$kHz in (\ref{fig:figure3b}e). These results confirm the directional wave beaming at high-order fold symmetries occurring at the identified anisotropic bands, also evidencing the twisting of the wave directionality for different frequencies.

\begin{figure}
	\centering
	{\includegraphics[scale=0.46]{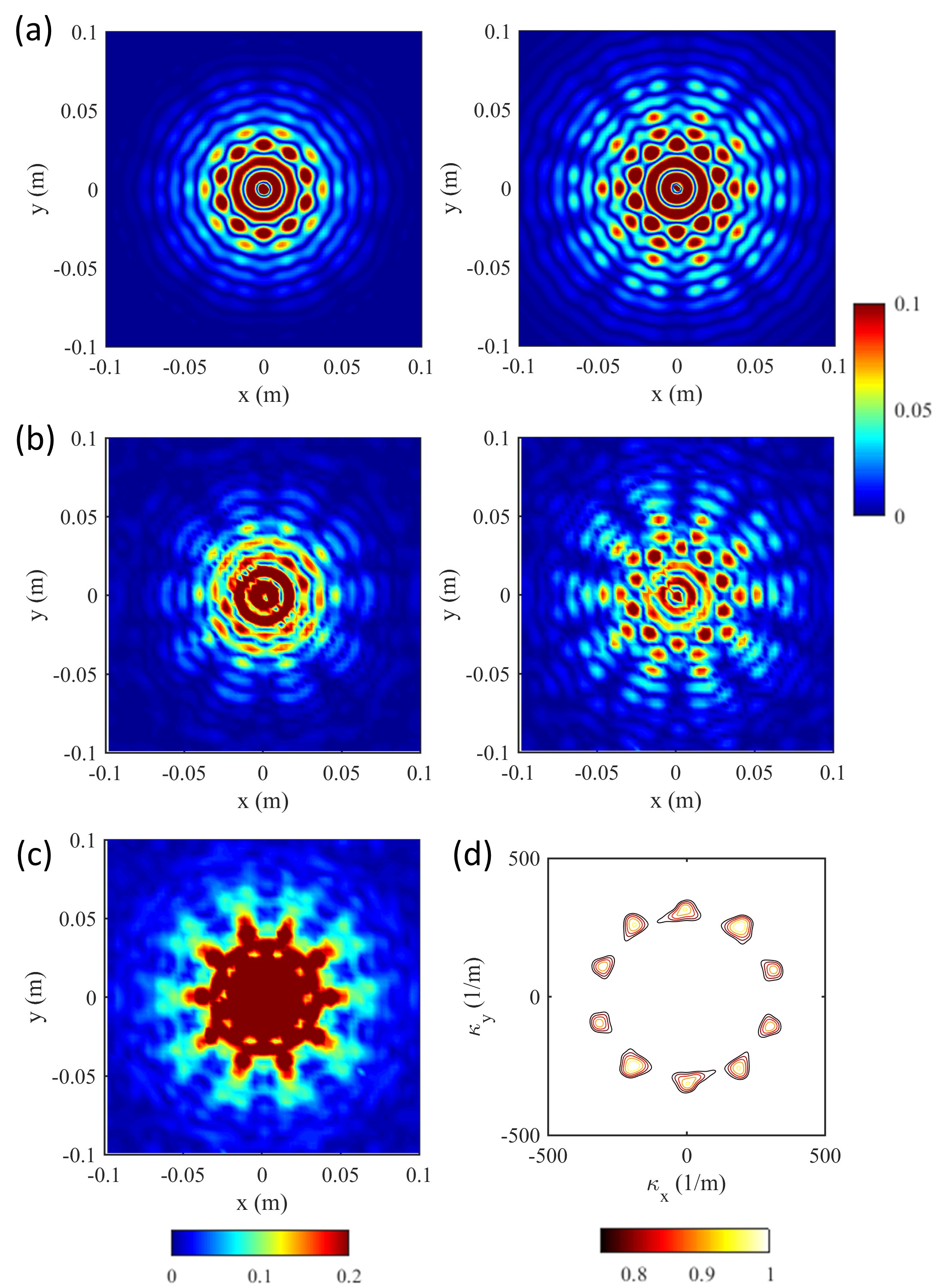}} 
	\caption{\label{fig:figure3} Wave beaming for the 10-fold quasicrystalline plate: numerical simulation at 30 kHz (a) and experimental observation at 24.8 kHz (b-d). A good agreement between simulation and experiment is achieved in the snapshots of the time response (a-b). Experimental RMS of the wave field averaging across all time (c) and RMS of the wave number contours averaging across all frequencies (d).} 
\end{figure}  

The numerically predicted 10-fold wave directionality around 30 kHz (Fig. \ref{fig:figure3b}(c)) is also confirmed experimentally in Fig. \ref{fig:figure3}. Despite a shift in frequency, a good agreement is observed between numerical and experimental wave-field results at the different time snapshots of Figs. \ref{fig:figure3}(a,b). A good agreement is also observed in the numerical and experimental RMS of the wave-fields (Fig. \ref{fig:figure3b}(c) and \ref{fig:figure3}(c), respectively). Both dynamic responses exhibit directionality along 10-fold symmetric directions, in agreement with the group velocity pattern depicted in Fig. \ref{fig:figure3b}(c). Finally, their RMS wave number contour obtained from the 3D-FT are also similar for both simulations (Fig. \ref{fig:figure3b}(c)) and experiments (Fig. \ref{fig:figure3}(d)).


\section{Wave Diffraction} 

We next illustrate how the directionalities provided by the higher order symmetries manifest in the wave diffraction. Diffraction occurs when a wave passes through an aperture or an obstacle, and has been largely employed in focusing, lensing and antennas \cite{Radi2017}. The combination of different apertures, i.e. metagratings, has been used to create specific wave fronts with required fold symmetry \cite{Popov2018}. However, wave branches with angles larger than $45^{\circ}$ in relation to the incident wave are usually difficult to be created. Diffraction in quasiperiodic configurations, for instance, can open new possibilities for superior directivity control in multiple angles as well as for loudspeakers with high quality perception \cite{Pasqual2010}. 

To illustrate the diffraction behavior, the upper half of the quasicrystal domain considered in the previous section ($N=10, x = [-0.1, \ 0.1]$ m and $y = [0, \ 0.2]$ m) is combined to an uniform plate ($x = [-0.1, \ 0.1]$ m and $y = [-0.075, \ 0]$ m) with constant thickness $h_{\tt U}$ = 4 mm (see SM for more details on the designs). Low reflection conditions are imposed at the boundaries to minimize backscattering. A line-source excitation is centered at the bottom of the uniform plate ($x = 0$ and $y = -0.075$ m), so that the incident wave propagates along the positive $y$ direction until it reaches the interface with the quasicrystalline domain ($y = 0$). The diffraction is illustrated by employing a sinusoidal burst excitation signal with 15 cycles for a center frequency of $9.3$kHz, corresponding to the first strong beaming behavior reported in Fig.~\ref{fig:figure3b}. 

\begin{figure*}
	\centering
	{\includegraphics[scale=0.525]{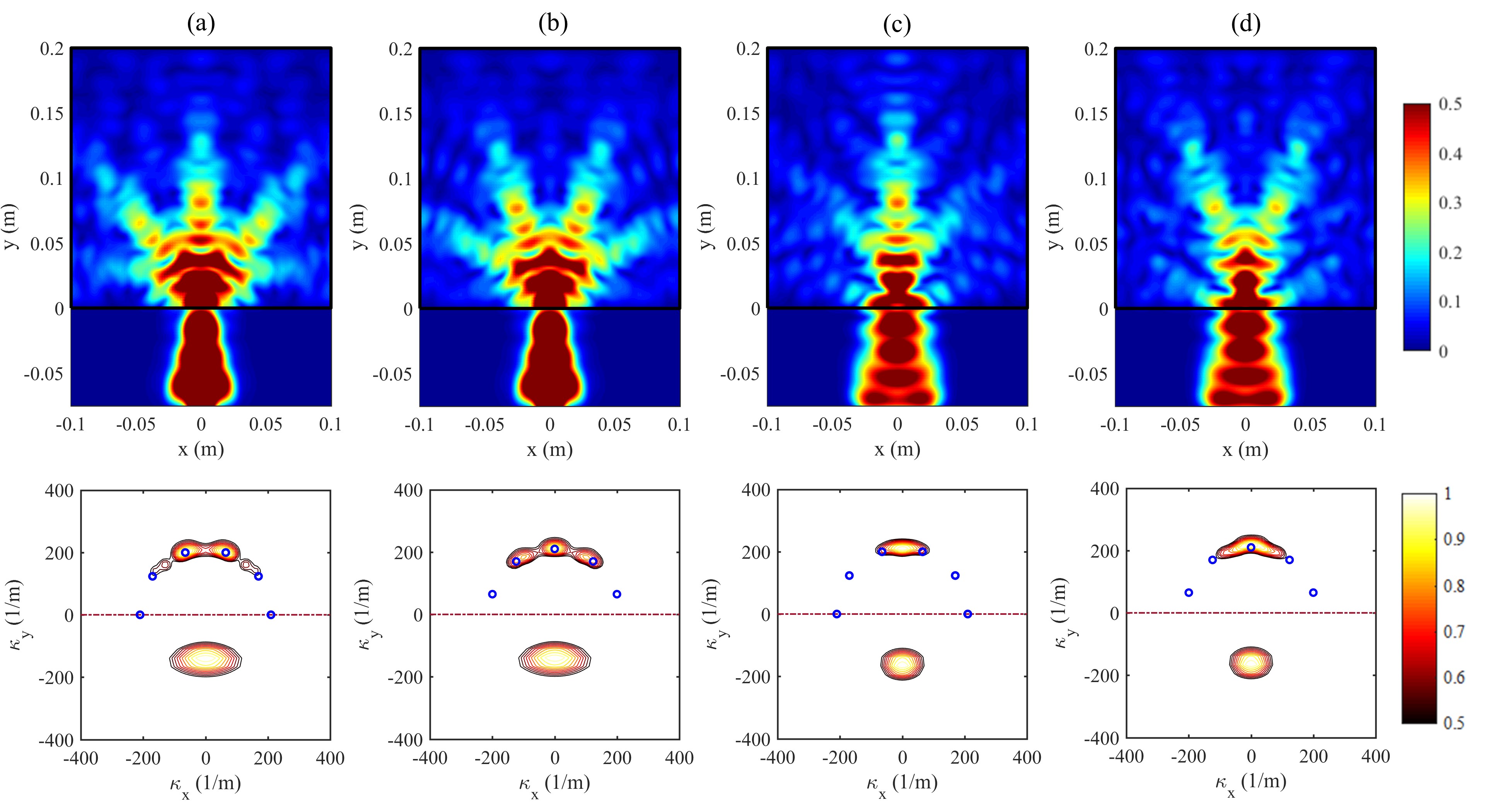}} 
	\caption{\label{fig:figure4} Wave diffraction at 9.3 kHz in a half of the extended 10-fold quasiperiodic metamaterial plate, where the incident wave with width $b_s = 0.02$ m (a-b) or $b_s = 0.06$ m (c-d) are aligned $\theta = 0$ (a, c) or twisted by $\theta = \theta_N/2$ (b, d) in relation to one of the  Bragg peaks in reciprocal space at 9.3 kHz. The first line depicts the RMS of the displacement field averaging across all time and the second line the correspondent RMS of the wave number contour averaging across all frequencies: $\kappa_y > 0$ for the quasicrystal plate and $\kappa_y < 0$ for the uniform plate. The blue circles (second line) correspond to the spectral content peaks of the 10-fold plate at 9.3 kHz, see Fig.~\ref{fig:figure3b} (a). Diffraction patterns with focusing from one to four branches have been created.} 
\end{figure*}

The results are summarized in Fig. \ref{fig:figure4} for different conditions that showcase different possible scenarios; the top row displays the RMS of the wave-field (a snapshot for each case is also shown in the SM), while the bottom row displays the RMS of the reciprocal space content (the upper half, $\kappa_y > 0$, corresponding to the waves propagating in the quasicrystal, and the bottom half, $\kappa_y < 0$, corresponding to the incident wave in the homogeneous plate). The results in (a,b) correspond to a narrow line source of width 20 mm, which provides a broad wave number content for $\kappa_x$ for the incident wave, while in (c,d) a wider line source of 60 mm produces an incident wave with narrower wave number content for $\kappa_x$. Also, in (a,c) the quasicrystalline plate is designed with the conventional procedure described in Section II, while in (b,d) the design peaks, and the plate symmetry, are twisted by $\theta_N/2=18^{o}$. These different conditions are selected to illustrate a wealth of possibilities for wave diffraction that result in different numbers and orientations of directional branches propagating in the quasicrystalline plate. 

The results are interpreted based on the intersection of the wave-number content of the incident wave (bottom half of FT panels), and the symmetry peaks of the dispersion at the excitation frequency (blue dots in upper half of the FT panels). In (a), the wave-number content of the incident wave reaches the zone between two Bragg peaks, which are mainly excited as evidenced by the 2D FTs. Three directional branches are observed to propagate in the plate and their directions are in agreement with the group velocity plots of Fig.~\ref{fig:figure3b} (a). For the same plate with narrower wave number content of the incident wave (c), only the central region between the same peaks gets excited, resulting in a single branch propagating into the quasicrystalline portion. When the plate is twisted (b,d), the high symmetry points of the dispersion (blue circles) and the associated directional branches also get twisted. The broader excitation in reciprocal space produces 4 propagating branches (b), while the narrower excitation produces only 2 (d), and their directions are accordingly twisted by $\theta_N/2=18^{o}$ with respect to those in the group velocity plot of Fig.~\ref{fig:figure3b} (a). These results highlight how multiple scenarios for wave diffraction can be envisioned by controlling the quasicrystal orientation and the source width, which can shift the wave behavior from beaming (i.e., multi-focal) to focusing (i.e., uni-focal).


\section{Conclusions} \label{sec:Conclusions}

The spatial design of phononic crystals and acoustic-elastic metamaterials is based on translational periodicity, and hence, their wave phenomena, such as the directionality, are restricted to the crystallographic symmetries (e.g., 2-, 4- and 6-fold). In this work, the wave beaming and diffraction were expanded to high-order rotational symmetries, such as 8- and 10-fold, by employing quasiperiodic elastic metamaterial plates. Their spectral contents were investigated using an approximate dispersion surface obtained from the time response and its Fourier transform. In some frequency zones, the wave behavior becomes highly anisotropic and, therefore, high-order wave directionality was observed on the dynamic response (e.g., beaming and diffraction). Experimental observations in practical prototypes were also performed, which facilitate the implementation in wave devices and structural components. This work provides powerful tools as well as highlights the rich underlying physics behind the wave phenomena in quasiperiodic architected materials. Moreover, it opens new possibilities for applications involving the unusual wave front directivity with high-order symmetry (e.g. 8-fold, 10-fold and so on), such as focusing, sensing and imaging beyond the symmetries provided by the periodic configurations.


\begin{acknowledgments}
D. Beli and C. De Marqui Jr. gratefully acknowledge the support from S\~{a}o Paulo Research Foundation (FAPESP) through grant reference numbers: 2018/18774-6, 2019/22464-5 and 2018/15894-0 (Research project - Periodic structure design and optimization for enhanced vibroacoustic performance: ENVIBRO). M. I. N. Rosa and M. Ruzzene gratefully acknowledge the support from the National Science Foundation (NSF) through the EFRI 1741685 grant and from the Army Research Office through grant W911NF-18-1-0036.
\end{acknowledgments}



\bibliography{references}

\begin{thebibliography}{61}%
\makeatletter
\providecommand \@ifxundefined [1]{%
 \@ifx{#1\undefined}
}%
\providecommand \@ifnum [1]{%
 \ifnum #1\expandafter \@firstoftwo
 \else \expandafter \@secondoftwo
 \fi
}%
\providecommand \@ifx [1]{%
 \ifx #1\expandafter \@firstoftwo
 \else \expandafter \@secondoftwo
 \fi
}%
\providecommand \natexlab [1]{#1}%
\providecommand \enquote  [1]{``#1''}%
\providecommand \bibnamefont  [1]{#1}%
\providecommand \bibfnamefont [1]{#1}%
\providecommand \citenamefont [1]{#1}%
\providecommand \href@noop [0]{\@secondoftwo}%
\providecommand \href [0]{\begingroup \@sanitize@url \@href}%
\providecommand \@href[1]{\@@startlink{#1}\@@href}%
\providecommand \@@href[1]{\endgroup#1\@@endlink}%
\providecommand \@sanitize@url [0]{\catcode `\\12\catcode `\$12\catcode
  `\&12\catcode `\#12\catcode `\^12\catcode `\_12\catcode `\%12\relax}%
\providecommand \@@startlink[1]{}%
\providecommand \@@endlink[0]{}%
\providecommand \url  [0]{\begingroup\@sanitize@url \@url }%
\providecommand \@url [1]{\endgroup\@href {#1}{\urlprefix }}%
\providecommand \urlprefix  [0]{URL }%
\providecommand \Eprint [0]{\href }%
\providecommand \doibase [0]{https://doi.org/}%
\providecommand \selectlanguage [0]{\@gobble}%
\providecommand \bibinfo  [0]{\@secondoftwo}%
\providecommand \bibfield  [0]{\@secondoftwo}%
\providecommand \translation [1]{[#1]}%
\providecommand \BibitemOpen [0]{}%
\providecommand \bibitemStop [0]{}%
\providecommand \bibitemNoStop [0]{.\EOS\space}%
\providecommand \EOS [0]{\spacefactor3000\relax}%
\providecommand \BibitemShut  [1]{\csname bibitem#1\endcsname}%
\let\auto@bib@innerbib\@empty
\bibitem [{\citenamefont {Joannopoulos}\ \emph {et~al.}(1997)\citenamefont
  {Joannopoulos}, \citenamefont {Villeneuve},\ and\ \citenamefont
  {Fan}}]{Joannopoulos1997}%
  \BibitemOpen
  \bibfield  {author} {\bibinfo {author} {\bibfnamefont {J.~D.}\ \bibnamefont
  {Joannopoulos}}, \bibinfo {author} {\bibfnamefont {P.~R.}\ \bibnamefont
  {Villeneuve}},\ and\ \bibinfo {author} {\bibfnamefont {S.}~\bibnamefont
  {Fan}},\ }\bibfield  {title} {\bibinfo {title} {Photonic crystals: putting a
  new twist on light},\ }\href {https://doi.org/10.1038/386143a0} {\bibfield
  {journal} {\bibinfo  {journal} {Nature}\ }\textbf {\bibinfo {volume} {386}},\
  \bibinfo {pages} {143} (\bibinfo {year} {1997})}\BibitemShut {NoStop}%
\bibitem [{\citenamefont {Hussein}\ \emph {et~al.}(2014)\citenamefont
  {Hussein}, \citenamefont {Leamy},\ and\ \citenamefont
  {Ruzzene}}]{Hussein2014}%
  \BibitemOpen
  \bibfield  {author} {\bibinfo {author} {\bibfnamefont {M.~I.}\ \bibnamefont
  {Hussein}}, \bibinfo {author} {\bibfnamefont {M.~J.}\ \bibnamefont {Leamy}},\
  and\ \bibinfo {author} {\bibfnamefont {M.}~\bibnamefont {Ruzzene}},\
  }\bibfield  {title} {\bibinfo {title} {Dynamics of phononic materials and
  structures: Historical origins, recent progress, and future outlook},\ }\href
  {https://doi.org/10.1115/1.4026911} {\bibfield  {journal} {\bibinfo
  {journal} {Applied Mechanics Reviews}\ }\textbf {\bibinfo {volume} {66}},\
  \bibinfo {pages} {040802} (\bibinfo {year} {2014})}\BibitemShut {NoStop}%
\bibitem [{\citenamefont {Christensen}\ \emph {et~al.}(2015)\citenamefont
  {Christensen}, \citenamefont {Kadic}, \citenamefont {Kraft},\ and\
  \citenamefont {Wegener}}]{Christensen2015}%
  \BibitemOpen
  \bibfield  {author} {\bibinfo {author} {\bibfnamefont {J.}~\bibnamefont
  {Christensen}}, \bibinfo {author} {\bibfnamefont {M.}~\bibnamefont {Kadic}},
  \bibinfo {author} {\bibfnamefont {O.}~\bibnamefont {Kraft}},\ and\ \bibinfo
  {author} {\bibfnamefont {M.}~\bibnamefont {Wegener}},\ }\bibfield  {title}
  {\bibinfo {title} {Vibrant times for mechanical metamaterials},\ }\href
  {https://doi.org/10.1557/mrc.2015.51} {\bibfield  {journal} {\bibinfo
  {journal} {MRS Communications}\ }\textbf {\bibinfo {volume} {5}},\ \bibinfo
  {pages} {453–462} (\bibinfo {year} {2015})}\BibitemShut {NoStop}%
\bibitem [{\citenamefont {Ma}\ and\ \citenamefont {Sheng}(2016)}]{Ma2016}%
  \BibitemOpen
  \bibfield  {author} {\bibinfo {author} {\bibfnamefont {G.}~\bibnamefont
  {Ma}}\ and\ \bibinfo {author} {\bibfnamefont {P.}~\bibnamefont {Sheng}},\
  }\bibfield  {title} {\bibinfo {title} {Acoustic metamaterials: From local
  resonances to broad horizons},\ }\href
  {https://doi.org/10.1126/sciadv.1501595} {\bibfield  {journal} {\bibinfo
  {journal} {Science Advances}\ }\textbf {\bibinfo {volume} {2}},\ \bibinfo
  {pages} {e1501595} (\bibinfo {year} {2016})}\BibitemShut {NoStop}%
\bibitem [{\citenamefont {Cummer}\ \emph {et~al.}(2016)\citenamefont {Cummer},
  \citenamefont {Christensen},\ and\ \citenamefont {Alù}}]{Cummer2016}%
  \BibitemOpen
  \bibfield  {author} {\bibinfo {author} {\bibfnamefont {S.~A.}\ \bibnamefont
  {Cummer}}, \bibinfo {author} {\bibfnamefont {J.}~\bibnamefont
  {Christensen}},\ and\ \bibinfo {author} {\bibfnamefont {A.}~\bibnamefont
  {Alù}},\ }\bibfield  {title} {\bibinfo {title} {Controlling sound with
  acoustic metamaterials},\ }\href {https://doi.org/10.1038/natrevmats.2016.1}
  {\bibfield  {journal} {\bibinfo  {journal} {Nature Reviews Materials}\
  }\textbf {\bibinfo {volume} {1}},\ \bibinfo {pages} {13} (\bibinfo {year}
  {2016})}\BibitemShut {NoStop}%
\bibitem [{\citenamefont {Assouar}\ \emph {et~al.}(2016)\citenamefont
  {Assouar}, \citenamefont {Liang}, \citenamefont {Wu}, \citenamefont {Li},
  \citenamefont {Cheng},\ and\ \citenamefont {Jing}}]{Assouar2018}%
  \BibitemOpen
  \bibfield  {author} {\bibinfo {author} {\bibfnamefont {B.}~\bibnamefont
  {Assouar}}, \bibinfo {author} {\bibfnamefont {B.}~\bibnamefont {Liang}},
  \bibinfo {author} {\bibfnamefont {Y.}~\bibnamefont {Wu}}, \bibinfo {author}
  {\bibfnamefont {Y.}~\bibnamefont {Li}}, \bibinfo {author} {\bibfnamefont
  {J.-C.}\ \bibnamefont {Cheng}},\ and\ \bibinfo {author} {\bibfnamefont
  {Y.}~\bibnamefont {Jing}},\ }\bibfield  {title} {\bibinfo {title}
  {Controlling sound with acoustic metamaterials},\ }\href
  {https://doi.org/https://doi.org/10.1038/s41578-018-0061-4} {\bibfield
  {journal} {\bibinfo  {journal} {Nature Reviews Materials}\ }\textbf {\bibinfo
  {volume} {3}},\ \bibinfo {pages} {460} (\bibinfo {year} {2016})}\BibitemShut
  {NoStop}%
\bibitem [{\citenamefont {Yang}\ \emph {et~al.}(2015)\citenamefont {Yang},
  \citenamefont {Gao}, \citenamefont {Shi}, \citenamefont {Lin}, \citenamefont
  {Gao}, \citenamefont {Chong},\ and\ \citenamefont
  {Zhang}}]{Yang2015TopAcoustics}%
  \BibitemOpen
  \bibfield  {author} {\bibinfo {author} {\bibfnamefont {Z.}~\bibnamefont
  {Yang}}, \bibinfo {author} {\bibfnamefont {F.}~\bibnamefont {Gao}}, \bibinfo
  {author} {\bibfnamefont {X.}~\bibnamefont {Shi}}, \bibinfo {author}
  {\bibfnamefont {X.}~\bibnamefont {Lin}}, \bibinfo {author} {\bibfnamefont
  {Z.}~\bibnamefont {Gao}}, \bibinfo {author} {\bibfnamefont {Y.}~\bibnamefont
  {Chong}},\ and\ \bibinfo {author} {\bibfnamefont {B.}~\bibnamefont {Zhang}},\
  }\bibfield  {title} {\bibinfo {title} {Topological acoustics},\ }\href
  {https://doi.org/10.1103/PhysRevLett.114.114301} {\bibfield  {journal}
  {\bibinfo  {journal} {Phys. Rev. Lett.}\ }\textbf {\bibinfo {volume} {114}},\
  \bibinfo {pages} {114301} (\bibinfo {year} {2015})}\BibitemShut {NoStop}%
\bibitem [{\citenamefont {Bandres}\ \emph {et~al.}(2016)\citenamefont
  {Bandres}, \citenamefont {Rechtsman},\ and\ \citenamefont
  {Segev}}]{Bandres2016}%
  \BibitemOpen
  \bibfield  {author} {\bibinfo {author} {\bibfnamefont {M.~A.}\ \bibnamefont
  {Bandres}}, \bibinfo {author} {\bibfnamefont {M.~C.}\ \bibnamefont
  {Rechtsman}},\ and\ \bibinfo {author} {\bibfnamefont {M.}~\bibnamefont
  {Segev}},\ }\bibfield  {title} {\bibinfo {title} {Topological photonic
  quasicrystals: Fractal topological spectrum and protected transport},\ }\href
  {https://doi.org/10.1103/PhysRevX.6.011016} {\bibfield  {journal} {\bibinfo
  {journal} {Phys. Rev. X}\ }\textbf {\bibinfo {volume} {6}},\ \bibinfo {pages}
  {011016} (\bibinfo {year} {2016})}\BibitemShut {NoStop}%
\bibitem [{\citenamefont {Fremling}\ \emph {et~al.}(2020)\citenamefont
  {Fremling}, \citenamefont {van Hooft}, \citenamefont {Smith},\ and\
  \citenamefont {Fritz}}]{Fremling2020}%
  \BibitemOpen
  \bibfield  {author} {\bibinfo {author} {\bibfnamefont {M.}~\bibnamefont
  {Fremling}}, \bibinfo {author} {\bibfnamefont {M.}~\bibnamefont {van Hooft}},
  \bibinfo {author} {\bibfnamefont {C.~M.}\ \bibnamefont {Smith}},\ and\
  \bibinfo {author} {\bibfnamefont {L.}~\bibnamefont {Fritz}},\ }\bibfield
  {title} {\bibinfo {title} {Existence of robust edge currents in
  sierpi\ifmmode \acute{n}\else \'{n}\fi{}ski fractals},\ }\href
  {https://doi.org/10.1103/PhysRevResearch.2.013044} {\bibfield  {journal}
  {\bibinfo  {journal} {Phys. Rev. Research}\ }\textbf {\bibinfo {volume}
  {2}},\ \bibinfo {pages} {013044} (\bibinfo {year} {2020})}\BibitemShut
  {NoStop}%
\bibitem [{\citenamefont {Celli}\ \emph {et~al.}(2019)\citenamefont {Celli},
  \citenamefont {Yousefzadeh}, \citenamefont {Daraio},\ and\ \citenamefont
  {Gonella}}]{Celli2019}%
  \BibitemOpen
  \bibfield  {author} {\bibinfo {author} {\bibfnamefont {P.}~\bibnamefont
  {Celli}}, \bibinfo {author} {\bibfnamefont {B.}~\bibnamefont {Yousefzadeh}},
  \bibinfo {author} {\bibfnamefont {C.}~\bibnamefont {Daraio}},\ and\ \bibinfo
  {author} {\bibfnamefont {S.}~\bibnamefont {Gonella}},\ }\bibfield  {title}
  {\bibinfo {title} {Bandgap widening by disorder in rainbow metamaterials},\
  }\href {https://doi.org/10.1063/1.5081916} {\bibfield  {journal} {\bibinfo
  {journal} {Applied Physics Letters}\ }\textbf {\bibinfo {volume} {114}},\
  \bibinfo {pages} {091903} (\bibinfo {year} {2019})}\BibitemShut {NoStop}%
\bibitem [{\citenamefont {Beli}\ \emph {et~al.}(2019)\citenamefont {Beli},
  \citenamefont {Fabro}, \citenamefont {Ruzzene},\ and\ \citenamefont
  {Arruda}}]{Beli2019}%
  \BibitemOpen
  \bibfield  {author} {\bibinfo {author} {\bibfnamefont {D.}~\bibnamefont
  {Beli}}, \bibinfo {author} {\bibfnamefont {A.~T.}\ \bibnamefont {Fabro}},
  \bibinfo {author} {\bibfnamefont {M.}~\bibnamefont {Ruzzene}},\ and\ \bibinfo
  {author} {\bibfnamefont {J.~R.~F.}\ \bibnamefont {Arruda}},\ }\bibfield
  {title} {\bibinfo {title} {Wave attenuation and trapping in 3d printed
  cantilever-in-mass metamaterials with spatially correlated variability},\
  }\href {https://doi.org/10.1038/s41598-019-41999-0} {\bibfield  {journal}
  {\bibinfo  {journal} {Scientific Reports}\ }\textbf {\bibinfo {volume} {9}},\
  \bibinfo {pages} {5617} (\bibinfo {year} {2019})}\BibitemShut {NoStop}%
\bibitem [{\citenamefont {Kraus}\ and\ \citenamefont
  {Zilberberg}(2016)}]{Kraus2016}%
  \BibitemOpen
  \bibfield  {author} {\bibinfo {author} {\bibfnamefont {Y.~E.}\ \bibnamefont
  {Kraus}}\ and\ \bibinfo {author} {\bibfnamefont {O.}~\bibnamefont
  {Zilberberg}},\ }\bibfield  {title} {\bibinfo {title} {Quasiperiodicity and
  topology transcend dimensions},\ }\href {https://doi.org/10.1038/nphys3784}
  {\bibfield  {journal} {\bibinfo  {journal} {Nature Physics}\ }\textbf
  {\bibinfo {volume} {12}},\ \bibinfo {pages} {624–626} (\bibinfo {year}
  {2016})}\BibitemShut {NoStop}%
\bibitem [{\citenamefont {Vardeny}\ \emph {et~al.}(2013)\citenamefont
  {Vardeny}, \citenamefont {Nahata},\ and\ \citenamefont
  {Agrawal}}]{Vardeny2013}%
  \BibitemOpen
  \bibfield  {author} {\bibinfo {author} {\bibfnamefont {Z.~V.}\ \bibnamefont
  {Vardeny}}, \bibinfo {author} {\bibfnamefont {A.}~\bibnamefont {Nahata}},\
  and\ \bibinfo {author} {\bibfnamefont {A.}~\bibnamefont {Agrawal}},\
  }\bibfield  {title} {\bibinfo {title} {Optics of photonic quasicrystals},\
  }\href {https://doi.org/10.1038/nphoton.2012.343} {\bibfield  {journal}
  {\bibinfo  {journal} {Nature Photonics}\ }\textbf {\bibinfo {volume} {7}},\
  \bibinfo {pages} {177} (\bibinfo {year} {2013})}\BibitemShut {NoStop}%
\bibitem [{\citenamefont {Shechtman}\ \emph {et~al.}(1984)\citenamefont
  {Shechtman}, \citenamefont {Blech}, \citenamefont {Gratias},\ and\
  \citenamefont {Cahn}}]{Shechtman1984}%
  \BibitemOpen
  \bibfield  {author} {\bibinfo {author} {\bibfnamefont {D.}~\bibnamefont
  {Shechtman}}, \bibinfo {author} {\bibfnamefont {I.}~\bibnamefont {Blech}},
  \bibinfo {author} {\bibfnamefont {D.}~\bibnamefont {Gratias}},\ and\ \bibinfo
  {author} {\bibfnamefont {J.~W.}\ \bibnamefont {Cahn}},\ }\bibfield  {title}
  {\bibinfo {title} {Metallic phase with long-range orientational order and no
  translational symmetry},\ }\href
  {https://doi.org/10.1103/PhysRevLett.53.1951} {\bibfield  {journal} {\bibinfo
   {journal} {Phys. Rev. Lett.}\ }\textbf {\bibinfo {volume} {53}},\ \bibinfo
  {pages} {1951} (\bibinfo {year} {1984})}\BibitemShut {NoStop}%
\bibitem [{\citenamefont {Levine}\ and\ \citenamefont
  {Steinhardt}(1984)}]{LevineSteinhardt1984}%
  \BibitemOpen
  \bibfield  {author} {\bibinfo {author} {\bibfnamefont {D.}~\bibnamefont
  {Levine}}\ and\ \bibinfo {author} {\bibfnamefont {P.~J.}\ \bibnamefont
  {Steinhardt}},\ }\bibfield  {title} {\bibinfo {title} {Quasicrystals: A new
  class of ordered structures},\ }\href
  {https://doi.org/10.1103/PhysRevLett.53.2477} {\bibfield  {journal} {\bibinfo
   {journal} {Phys. Rev. Lett.}\ }\textbf {\bibinfo {volume} {53}},\ \bibinfo
  {pages} {2477} (\bibinfo {year} {1984})}\BibitemShut {NoStop}%
\bibitem [{\citenamefont {Notomi}\ \emph {et~al.}(2004)\citenamefont {Notomi},
  \citenamefont {Suzuki}, \citenamefont {Tamamura},\ and\ \citenamefont
  {Edagawa}}]{Notomi2004}%
  \BibitemOpen
  \bibfield  {author} {\bibinfo {author} {\bibfnamefont {M.}~\bibnamefont
  {Notomi}}, \bibinfo {author} {\bibfnamefont {H.}~\bibnamefont {Suzuki}},
  \bibinfo {author} {\bibfnamefont {T.}~\bibnamefont {Tamamura}},\ and\
  \bibinfo {author} {\bibfnamefont {K.}~\bibnamefont {Edagawa}},\ }\bibfield
  {title} {\bibinfo {title} {Lasing action due to the two-dimensional
  quasiperiodicity of photonic quasicrystals with a penrose lattice},\ }\href
  {https://doi.org/10.1103/PhysRevLett.92.123906} {\bibfield  {journal}
  {\bibinfo  {journal} {Phys. Rev. Lett.}\ }\textbf {\bibinfo {volume} {92}},\
  \bibinfo {pages} {123906} (\bibinfo {year} {2004})}\BibitemShut {NoStop}%
\bibitem [{\citenamefont {Luo}\ \emph {et~al.}(2012)\citenamefont {Luo},
  \citenamefont {Du}, \citenamefont {Dai}, \citenamefont {Demir}, \citenamefont
  {Yang}, \citenamefont {Ji},\ and\ \citenamefont {Sun}}]{Luo2012}%
  \BibitemOpen
  \bibfield  {author} {\bibinfo {author} {\bibfnamefont {D.}~\bibnamefont
  {Luo}}, \bibinfo {author} {\bibfnamefont {Q.~G.}\ \bibnamefont {Du}},
  \bibinfo {author} {\bibfnamefont {H.~T.}\ \bibnamefont {Dai}}, \bibinfo
  {author} {\bibfnamefont {H.~V.}\ \bibnamefont {Demir}}, \bibinfo {author}
  {\bibfnamefont {H.~Z.}\ \bibnamefont {Yang}}, \bibinfo {author}
  {\bibfnamefont {W.}~\bibnamefont {Ji}},\ and\ \bibinfo {author}
  {\bibfnamefont {X.~W.}\ \bibnamefont {Sun}},\ }\bibfield  {title} {\bibinfo
  {title} {Strongly linearly polarized low threshold lasing of all organic
  photonic quasicrystals},\ }\href {https://doi.org/10.1038/srep00627}
  {\bibfield  {journal} {\bibinfo  {journal} {Scientific Reports}\ }\textbf
  {\bibinfo {volume} {2}},\ \bibinfo {pages} {627} (\bibinfo {year}
  {2012})}\BibitemShut {NoStop}%
\bibitem [{\citenamefont {Han}\ \emph {et~al.}(2020)\citenamefont {Han},
  \citenamefont {Xie}, \citenamefont {Liu},\ and\ \citenamefont
  {Liu}}]{Han2020}%
  \BibitemOpen
  \bibfield  {author} {\bibinfo {author} {\bibfnamefont {J.}~\bibnamefont
  {Han}}, \bibinfo {author} {\bibfnamefont {J.}~\bibnamefont {Xie}}, \bibinfo
  {author} {\bibfnamefont {E.}~\bibnamefont {Liu}},\ and\ \bibinfo {author}
  {\bibfnamefont {J.}~\bibnamefont {Liu}},\ }\bibfield  {title} {\bibinfo
  {title} {Super-resolution imaging in multi-broadband of a ten-fold
  penrose-type phononic quasi-crystal flat lens},\ }\href
  {https://doi.org/https://doi.org/10.1016/j.rinp.2020.103418} {\bibfield
  {journal} {\bibinfo  {journal} {Results in Physics}\ }\textbf {\bibinfo
  {volume} {19}},\ \bibinfo {pages} {103418} (\bibinfo {year}
  {2020})}\BibitemShut {NoStop}%
\bibitem [{\citenamefont {An}\ \emph {et~al.}(2020)\citenamefont {An},
  \citenamefont {Gao},\ and\ \citenamefont {Ouyang}}]{An2020}%
  \BibitemOpen
  \bibfield  {author} {\bibinfo {author} {\bibfnamefont {Y.}~\bibnamefont
  {An}}, \bibinfo {author} {\bibfnamefont {Z.}~\bibnamefont {Gao}},\ and\
  \bibinfo {author} {\bibfnamefont {Z.}~\bibnamefont {Ouyang}},\ }\bibfield
  {title} {\bibinfo {title} {Surface wave photonic quasicrystal},\ }\href
  {https://doi.org/10.1063/1.5139267} {\bibfield  {journal} {\bibinfo
  {journal} {Applied Physics Letters}\ }\textbf {\bibinfo {volume} {116}},\
  \bibinfo {pages} {151104} (\bibinfo {year} {2020})}\BibitemShut {NoStop}%
\bibitem [{\citenamefont {Sakai}\ and\ \citenamefont
  {Arita}(2019)}]{Sakai2019}%
  \BibitemOpen
  \bibfield  {author} {\bibinfo {author} {\bibfnamefont {S.}~\bibnamefont
  {Sakai}}\ and\ \bibinfo {author} {\bibfnamefont {R.}~\bibnamefont {Arita}},\
  }\bibfield  {title} {\bibinfo {title} {Exotic pairing state in
  quasicrystalline superconductors under a magnetic field},\ }\href
  {https://doi.org/10.1103/PhysRevResearch.1.022002} {\bibfield  {journal}
  {\bibinfo  {journal} {Phys. Rev. Research}\ }\textbf {\bibinfo {volume}
  {1}},\ \bibinfo {pages} {022002} (\bibinfo {year} {2019})}\BibitemShut
  {NoStop}%
\bibitem [{\citenamefont {Apigo}\ \emph {et~al.}(2019)\citenamefont {Apigo},
  \citenamefont {Cheng}, \citenamefont {Dobiszewski}, \citenamefont {Prodan},\
  and\ \citenamefont {Prodan}}]{apigo2019observation}%
  \BibitemOpen
  \bibfield  {author} {\bibinfo {author} {\bibfnamefont {D.~J.}\ \bibnamefont
  {Apigo}}, \bibinfo {author} {\bibfnamefont {W.}~\bibnamefont {Cheng}},
  \bibinfo {author} {\bibfnamefont {K.~F.}\ \bibnamefont {Dobiszewski}},
  \bibinfo {author} {\bibfnamefont {E.}~\bibnamefont {Prodan}},\ and\ \bibinfo
  {author} {\bibfnamefont {C.}~\bibnamefont {Prodan}},\ }\bibfield  {title}
  {\bibinfo {title} {Observation of topological edge modes in a quasiperiodic
  acoustic waveguide},\ }\href {https://doi.org/10.1103/PhysRevLett.122.095501}
  {\bibfield  {journal} {\bibinfo  {journal} {Physical review letters}\
  }\textbf {\bibinfo {volume} {122}},\ \bibinfo {pages} {095501} (\bibinfo
  {year} {2019})}\BibitemShut {NoStop}%
\bibitem [{\citenamefont {Ni}\ \emph {et~al.}(2019)\citenamefont {Ni},
  \citenamefont {Chen}, \citenamefont {Weiner}, \citenamefont {Apigo},
  \citenamefont {Prodan}, \citenamefont {Al{\`u}}, \citenamefont {Prodan},\
  and\ \citenamefont {Khanikaev}}]{ni2019observation}%
  \BibitemOpen
  \bibfield  {author} {\bibinfo {author} {\bibfnamefont {X.}~\bibnamefont
  {Ni}}, \bibinfo {author} {\bibfnamefont {K.}~\bibnamefont {Chen}}, \bibinfo
  {author} {\bibfnamefont {M.}~\bibnamefont {Weiner}}, \bibinfo {author}
  {\bibfnamefont {D.~J.}\ \bibnamefont {Apigo}}, \bibinfo {author}
  {\bibfnamefont {C.}~\bibnamefont {Prodan}}, \bibinfo {author} {\bibfnamefont
  {A.}~\bibnamefont {Al{\`u}}}, \bibinfo {author} {\bibfnamefont
  {E.}~\bibnamefont {Prodan}},\ and\ \bibinfo {author} {\bibfnamefont {A.~B.}\
  \bibnamefont {Khanikaev}},\ }\bibfield  {title} {\bibinfo {title}
  {Observation of hofstadter butterfly and topological edge states in
  reconfigurable quasi-periodic acoustic crystals},\ }\href
  {https://doi.org/10.1038/s42005-019-0151-7} {\bibfield  {journal} {\bibinfo
  {journal} {Communications Physics}\ }\textbf {\bibinfo {volume} {2}},\
  \bibinfo {pages} {1} (\bibinfo {year} {2019})}\BibitemShut {NoStop}%
\bibitem [{\citenamefont {Wang}\ and\ \citenamefont
  {Sigmund}(2020)}]{Wang2020}%
  \BibitemOpen
  \bibfield  {author} {\bibinfo {author} {\bibfnamefont {Y.}~\bibnamefont
  {Wang}}\ and\ \bibinfo {author} {\bibfnamefont {O.}~\bibnamefont {Sigmund}},\
  }\bibfield  {title} {\bibinfo {title} {Quasiperiodic mechanical metamaterials
  with extreme isotropic stiffness},\ }\href
  {https://doi.org/https://doi.org/10.1016/j.eml.2019.100596} {\bibfield
  {journal} {\bibinfo  {journal} {Extreme Mechanics Letters}\ }\textbf
  {\bibinfo {volume} {34}},\ \bibinfo {pages} {100596} (\bibinfo {year}
  {2020})}\BibitemShut {NoStop}%
\bibitem [{\citenamefont {Chen}\ \emph
  {et~al.}(2020{\natexlab{a}})\citenamefont {Chen}, \citenamefont {Kadic},
  \citenamefont {Guenneau},\ and\ \citenamefont {Wegener}}]{Chen2020}%
  \BibitemOpen
  \bibfield  {author} {\bibinfo {author} {\bibfnamefont {Y.}~\bibnamefont
  {Chen}}, \bibinfo {author} {\bibfnamefont {M.}~\bibnamefont {Kadic}},
  \bibinfo {author} {\bibfnamefont {S.}~\bibnamefont {Guenneau}},\ and\
  \bibinfo {author} {\bibfnamefont {M.}~\bibnamefont {Wegener}},\ }\bibfield
  {title} {\bibinfo {title} {Isotropic chiral acoustic phonons in 3d
  quasicrystalline metamaterials},\ }\href
  {https://doi.org/10.1103/PhysRevLett.124.235502} {\bibfield  {journal}
  {\bibinfo  {journal} {Phys. Rev. Lett.}\ }\textbf {\bibinfo {volume} {124}},\
  \bibinfo {pages} {235502} (\bibinfo {year} {2020}{\natexlab{a}})}\BibitemShut
  {NoStop}%
\bibitem [{\citenamefont {Beli}\ \emph {et~al.}(2021)\citenamefont {Beli},
  \citenamefont {Rosa}, \citenamefont {Marqui},\ and\ \citenamefont
  {Ruzzene}}]{Beli2021}%
  \BibitemOpen
  \bibfield  {author} {\bibinfo {author} {\bibfnamefont {D.}~\bibnamefont
  {Beli}}, \bibinfo {author} {\bibfnamefont {M.~I.~N.}\ \bibnamefont {Rosa}},
  \bibinfo {author} {\bibfnamefont {C.~D.}\ \bibnamefont {Marqui}},\ and\
  \bibinfo {author} {\bibfnamefont {M.}~\bibnamefont {Ruzzene}},\ }\bibfield
  {title} {\bibinfo {title} {Mechanics and dynamics of two-dimensional
  quasicrystalline composites},\ }\href
  {https://doi.org/https://doi.org/10.1016/j.eml.2021.101220} {\bibfield
  {journal} {\bibinfo  {journal} {Extreme Mechanics Letters}\ }\textbf
  {\bibinfo {volume} {44}},\ \bibinfo {pages} {101220} (\bibinfo {year}
  {2021})}\BibitemShut {NoStop}%
\bibitem [{\citenamefont {Chen}\ \emph {et~al.}(2021)\citenamefont {Chen},
  \citenamefont {Frenzel}, \citenamefont {Zhang}, \citenamefont {Kadic},\ and\
  \citenamefont {Wegener}}]{Chen2021}%
  \BibitemOpen
  \bibfield  {author} {\bibinfo {author} {\bibfnamefont {Y.}~\bibnamefont
  {Chen}}, \bibinfo {author} {\bibfnamefont {T.}~\bibnamefont {Frenzel}},
  \bibinfo {author} {\bibfnamefont {Q.}~\bibnamefont {Zhang}}, \bibinfo
  {author} {\bibfnamefont {M.}~\bibnamefont {Kadic}},\ and\ \bibinfo {author}
  {\bibfnamefont {M.}~\bibnamefont {Wegener}},\ }\bibfield  {title} {\bibinfo
  {title} {Cubic metamaterial crystal supporting broadband isotropic chiral
  phonons},\ }\href {https://doi.org/10.1103/PhysRevMaterials.5.025201}
  {\bibfield  {journal} {\bibinfo  {journal} {Phys. Rev. Materials}\ }\textbf
  {\bibinfo {volume} {5}},\ \bibinfo {pages} {025201} (\bibinfo {year}
  {2021})}\BibitemShut {NoStop}%
\bibitem [{\citenamefont {Florescu}\ \emph {et~al.}(2009)\citenamefont
  {Florescu}, \citenamefont {Torquato},\ and\ \citenamefont
  {Steinhardt}}]{Florescu2009}%
  \BibitemOpen
  \bibfield  {author} {\bibinfo {author} {\bibfnamefont {M.}~\bibnamefont
  {Florescu}}, \bibinfo {author} {\bibfnamefont {S.}~\bibnamefont {Torquato}},\
  and\ \bibinfo {author} {\bibfnamefont {P.~J.}\ \bibnamefont {Steinhardt}},\
  }\bibfield  {title} {\bibinfo {title} {Complete band gaps in two-dimensional
  photonic quasicrystals},\ }\href {https://doi.org/10.1103/PhysRevB.80.155112}
  {\bibfield  {journal} {\bibinfo  {journal} {Phys. Rev. B}\ }\textbf {\bibinfo
  {volume} {80}},\ \bibinfo {pages} {155112} (\bibinfo {year}
  {2009})}\BibitemShut {NoStop}%
\bibitem [{\citenamefont {Kaliteevski}\ \emph {et~al.}(2000)\citenamefont
  {Kaliteevski}, \citenamefont {Brand}, \citenamefont {Abram}, \citenamefont
  {Krauss}, \citenamefont {Rue},\ and\ \citenamefont
  {Millar}}]{Kaliteevski2000}%
  \BibitemOpen
  \bibfield  {author} {\bibinfo {author} {\bibfnamefont {M.~A.}\ \bibnamefont
  {Kaliteevski}}, \bibinfo {author} {\bibfnamefont {S.}~\bibnamefont {Brand}},
  \bibinfo {author} {\bibfnamefont {R.~A.}\ \bibnamefont {Abram}}, \bibinfo
  {author} {\bibfnamefont {T.~F.}\ \bibnamefont {Krauss}}, \bibinfo {author}
  {\bibfnamefont {R.~D.}\ \bibnamefont {Rue}},\ and\ \bibinfo {author}
  {\bibfnamefont {P.}~\bibnamefont {Millar}},\ }\bibfield  {title} {\bibinfo
  {title} {Two-dimensional penrose-tiled photonic quasicrystals: from
  diffraction pattern to band structure},\ }\href
  {https://doi.org/10.1088/0957-4484/11/4/316} {\bibfield  {journal} {\bibinfo
  {journal} {Nanotechnology}\ }\textbf {\bibinfo {volume} {11}},\ \bibinfo
  {pages} {274} (\bibinfo {year} {2000})}\BibitemShut {NoStop}%
\bibitem [{\citenamefont {Rotenberg}\ \emph {et~al.}(2000)\citenamefont
  {Rotenberg}, \citenamefont {Theis}, \citenamefont {Horn},\ and\ \citenamefont
  {Gille}}]{Rotenberg2000}%
  \BibitemOpen
  \bibfield  {author} {\bibinfo {author} {\bibfnamefont {E.}~\bibnamefont
  {Rotenberg}}, \bibinfo {author} {\bibfnamefont {W.}~\bibnamefont {Theis}},
  \bibinfo {author} {\bibfnamefont {K.}~\bibnamefont {Horn}},\ and\ \bibinfo
  {author} {\bibfnamefont {P.}~\bibnamefont {Gille}},\ }\bibfield  {title}
  {\bibinfo {title} {Quasicrystalline valence bands in decagonal alnico},\
  }\href {https://doi.org/10.1038/35020519} {\bibfield  {journal} {\bibinfo
  {journal} {Nature}\ }\textbf {\bibinfo {volume} {406}},\ \bibinfo {pages}
  {602} (\bibinfo {year} {2000})}\BibitemShut {NoStop}%
\bibitem [{\citenamefont {Gambaudo}\ and\ \citenamefont
  {Vignolo}(2014)}]{Gambaudo2014}%
  \BibitemOpen
  \bibfield  {author} {\bibinfo {author} {\bibfnamefont {J.-M.}\ \bibnamefont
  {Gambaudo}}\ and\ \bibinfo {author} {\bibfnamefont {P.}~\bibnamefont
  {Vignolo}},\ }\bibfield  {title} {\bibinfo {title} {Brillouin zone labelling
  for quasicrystals},\ }\href {https://doi.org/10.1088/1367-2630/16/4/043013}
  {\bibfield  {journal} {\bibinfo  {journal} {New Journal of Physics}\ }\textbf
  {\bibinfo {volume} {16}},\ \bibinfo {pages} {043013} (\bibinfo {year}
  {2014})}\BibitemShut {NoStop}%
\bibitem [{\citenamefont {Fuchs}\ \emph {et~al.}(2018)\citenamefont {Fuchs},
  \citenamefont {Mosseri},\ and\ \citenamefont {Vidal}}]{Fuchs2018}%
  \BibitemOpen
  \bibfield  {author} {\bibinfo {author} {\bibfnamefont {J.-N.}\ \bibnamefont
  {Fuchs}}, \bibinfo {author} {\bibfnamefont {R.}~\bibnamefont {Mosseri}},\
  and\ \bibinfo {author} {\bibfnamefont {J.}~\bibnamefont {Vidal}},\ }\bibfield
   {title} {\bibinfo {title} {Landau levels in quasicrystals},\ }\href
  {https://doi.org/10.1103/PhysRevB.98.165427} {\bibfield  {journal} {\bibinfo
  {journal} {Phys. Rev. B}\ }\textbf {\bibinfo {volume} {98}},\ \bibinfo
  {pages} {165427} (\bibinfo {year} {2018})}\BibitemShut {NoStop}%
\bibitem [{\citenamefont {Cheng}\ \emph {et~al.}(1999)\citenamefont {Cheng},
  \citenamefont {Li}, \citenamefont {Chan},\ and\ \citenamefont
  {Zhang}}]{Cheng1999}%
  \BibitemOpen
  \bibfield  {author} {\bibinfo {author} {\bibfnamefont {S.~S.~M.}\
  \bibnamefont {Cheng}}, \bibinfo {author} {\bibfnamefont {L.-M.}\ \bibnamefont
  {Li}}, \bibinfo {author} {\bibfnamefont {C.~T.}\ \bibnamefont {Chan}},\ and\
  \bibinfo {author} {\bibfnamefont {Z.~Q.}\ \bibnamefont {Zhang}},\ }\bibfield
  {title} {\bibinfo {title} {Defect and transmission properties of
  two-dimensional quasiperiodic photonic band-gap systems},\ }\href
  {https://doi.org/10.1103/PhysRevB.59.4091} {\bibfield  {journal} {\bibinfo
  {journal} {Phys. Rev. B}\ }\textbf {\bibinfo {volume} {59}},\ \bibinfo
  {pages} {4091} (\bibinfo {year} {1999})}\BibitemShut {NoStop}%
\bibitem [{\citenamefont {Wang}\ \emph {et~al.}(2006)\citenamefont {Wang},
  \citenamefont {Liu}, \citenamefont {Zhang}, \citenamefont {Feng},\ and\
  \citenamefont {Li}}]{Wang2006}%
  \BibitemOpen
  \bibfield  {author} {\bibinfo {author} {\bibfnamefont {Y.}~\bibnamefont
  {Wang}}, \bibinfo {author} {\bibfnamefont {J.}~\bibnamefont {Liu}}, \bibinfo
  {author} {\bibfnamefont {B.}~\bibnamefont {Zhang}}, \bibinfo {author}
  {\bibfnamefont {S.}~\bibnamefont {Feng}},\ and\ \bibinfo {author}
  {\bibfnamefont {Z.-Y.}\ \bibnamefont {Li}},\ }\bibfield  {title} {\bibinfo
  {title} {Simulations of defect-free coupled-resonator optical waveguides
  constructed in 12-fold quasiperiodic photonic crystals},\ }\href
  {https://doi.org/10.1103/PhysRevB.73.155107} {\bibfield  {journal} {\bibinfo
  {journal} {Phys. Rev. B}\ }\textbf {\bibinfo {volume} {73}},\ \bibinfo
  {pages} {155107} (\bibinfo {year} {2006})}\BibitemShut {NoStop}%
\bibitem [{\citenamefont {Sinelnik}\ \emph {et~al.}(2020)\citenamefont
  {Sinelnik}, \citenamefont {Shishkin}, \citenamefont {Yu}, \citenamefont
  {Samusev}, \citenamefont {Belov}, \citenamefont {Limonov}, \citenamefont
  {Ginzburg},\ and\ \citenamefont {Rybin}}]{Sinelnik2020}%
  \BibitemOpen
  \bibfield  {author} {\bibinfo {author} {\bibfnamefont {A.~D.}\ \bibnamefont
  {Sinelnik}}, \bibinfo {author} {\bibfnamefont {I.~I.}\ \bibnamefont
  {Shishkin}}, \bibinfo {author} {\bibfnamefont {X.}~\bibnamefont {Yu}},
  \bibinfo {author} {\bibfnamefont {K.~B.}\ \bibnamefont {Samusev}}, \bibinfo
  {author} {\bibfnamefont {P.~A.}\ \bibnamefont {Belov}}, \bibinfo {author}
  {\bibfnamefont {M.~F.}\ \bibnamefont {Limonov}}, \bibinfo {author}
  {\bibfnamefont {P.}~\bibnamefont {Ginzburg}},\ and\ \bibinfo {author}
  {\bibfnamefont {M.~V.}\ \bibnamefont {Rybin}},\ }\bibfield  {title} {\bibinfo
  {title} {Experimental observation of intrinsic light localization in photonic
  icosahedral quasicrystals},\ }\href
  {https://doi.org/https://doi.org/10.1002/adom.202001170} {\bibfield
  {journal} {\bibinfo  {journal} {Advanced Optical Materials}\ }\textbf
  {\bibinfo {volume} {8}},\ \bibinfo {pages} {2001170} (\bibinfo {year}
  {2020})}\BibitemShut {NoStop}%
\bibitem [{\citenamefont {Fuchs}\ and\ \citenamefont
  {Vidal}(2016)}]{Fuchs2016Top}%
  \BibitemOpen
  \bibfield  {author} {\bibinfo {author} {\bibfnamefont {J.-N.}\ \bibnamefont
  {Fuchs}}\ and\ \bibinfo {author} {\bibfnamefont {J.}~\bibnamefont {Vidal}},\
  }\bibfield  {title} {\bibinfo {title} {Hofstadter butterfly of a
  quasicrystal},\ }\href {https://doi.org/10.1103/PhysRevB.94.205437}
  {\bibfield  {journal} {\bibinfo  {journal} {Phys. Rev. B}\ }\textbf {\bibinfo
  {volume} {94}},\ \bibinfo {pages} {205437} (\bibinfo {year}
  {2016})}\BibitemShut {NoStop}%
\bibitem [{\citenamefont {Apigo}\ \emph {et~al.}(2018)\citenamefont {Apigo},
  \citenamefont {Qian}, \citenamefont {Prodan},\ and\ \citenamefont
  {Prodan}}]{apigo2018topological}%
  \BibitemOpen
  \bibfield  {author} {\bibinfo {author} {\bibfnamefont {D.~J.}\ \bibnamefont
  {Apigo}}, \bibinfo {author} {\bibfnamefont {K.}~\bibnamefont {Qian}},
  \bibinfo {author} {\bibfnamefont {C.}~\bibnamefont {Prodan}},\ and\ \bibinfo
  {author} {\bibfnamefont {E.}~\bibnamefont {Prodan}},\ }\bibfield  {title}
  {\bibinfo {title} {Topological edge modes by smart patterning},\ }\href
  {https://doi.org/10.1103/PhysRevMaterials.2.124203} {\bibfield  {journal}
  {\bibinfo  {journal} {Physical Review Materials}\ }\textbf {\bibinfo {volume}
  {2}},\ \bibinfo {pages} {124203} (\bibinfo {year} {2018})}\BibitemShut
  {NoStop}%
\bibitem [{\citenamefont {Rosa}\ \emph {et~al.}(2019)\citenamefont {Rosa},
  \citenamefont {Pal}, \citenamefont {Arruda},\ and\ \citenamefont
  {Ruzzene}}]{Rosa2019}%
  \BibitemOpen
  \bibfield  {author} {\bibinfo {author} {\bibfnamefont {M.~I.~N.}\
  \bibnamefont {Rosa}}, \bibinfo {author} {\bibfnamefont {R.~K.}\ \bibnamefont
  {Pal}}, \bibinfo {author} {\bibfnamefont {J.~R.~F.}\ \bibnamefont {Arruda}},\
  and\ \bibinfo {author} {\bibfnamefont {M.}~\bibnamefont {Ruzzene}},\
  }\bibfield  {title} {\bibinfo {title} {Edge states and topological pumping in
  spatially modulated elastic lattices},\ }\href
  {https://doi.org/10.1103/PhysRevLett.123.034301} {\bibfield  {journal}
  {\bibinfo  {journal} {Phys. Rev. Lett.}\ }\textbf {\bibinfo {volume} {123}},\
  \bibinfo {pages} {034301} (\bibinfo {year} {2019})}\BibitemShut {NoStop}%
\bibitem [{\citenamefont {Zhou}\ \emph {et~al.}(2019)\citenamefont {Zhou},
  \citenamefont {Zhang},\ and\ \citenamefont {Mao}}]{zhou2019topological}%
  \BibitemOpen
  \bibfield  {author} {\bibinfo {author} {\bibfnamefont {D.}~\bibnamefont
  {Zhou}}, \bibinfo {author} {\bibfnamefont {L.}~\bibnamefont {Zhang}},\ and\
  \bibinfo {author} {\bibfnamefont {X.}~\bibnamefont {Mao}},\ }\bibfield
  {title} {\bibinfo {title} {Topological boundary floppy modes in
  quasicrystals},\ }\href@noop {} {\bibfield  {journal} {\bibinfo  {journal}
  {Physical Review X}\ }\textbf {\bibinfo {volume} {9}},\ \bibinfo {pages}
  {021054} (\bibinfo {year} {2019})}\BibitemShut {NoStop}%
\bibitem [{\citenamefont {Chen}\ \emph
  {et~al.}(2020{\natexlab{b}})\citenamefont {Chen}, \citenamefont {Chen},
  \citenamefont {Gao}, \citenamefont {Zhou},\ and\ \citenamefont
  {Xu}}]{Chen2020Top}%
  \BibitemOpen
  \bibfield  {author} {\bibinfo {author} {\bibfnamefont {R.}~\bibnamefont
  {Chen}}, \bibinfo {author} {\bibfnamefont {C.-Z.}\ \bibnamefont {Chen}},
  \bibinfo {author} {\bibfnamefont {J.-H.}\ \bibnamefont {Gao}}, \bibinfo
  {author} {\bibfnamefont {B.}~\bibnamefont {Zhou}},\ and\ \bibinfo {author}
  {\bibfnamefont {D.-H.}\ \bibnamefont {Xu}},\ }\bibfield  {title} {\bibinfo
  {title} {Higher-order topological insulators in quasicrystals},\ }\href
  {https://doi.org/10.1103/PhysRevLett.124.036803} {\bibfield  {journal}
  {\bibinfo  {journal} {Phys. Rev. Lett.}\ }\textbf {\bibinfo {volume} {124}},\
  \bibinfo {pages} {036803} (\bibinfo {year} {2020}{\natexlab{b}})}\BibitemShut
  {NoStop}%
\bibitem [{\citenamefont {Cheng}\ \emph {et~al.}(2020)\citenamefont {Cheng},
  \citenamefont {Prodan},\ and\ \citenamefont
  {Prodan}}]{cheng2020experimental}%
  \BibitemOpen
  \bibfield  {author} {\bibinfo {author} {\bibfnamefont {W.}~\bibnamefont
  {Cheng}}, \bibinfo {author} {\bibfnamefont {E.}~\bibnamefont {Prodan}},\ and\
  \bibinfo {author} {\bibfnamefont {C.}~\bibnamefont {Prodan}},\ }\bibfield
  {title} {\bibinfo {title} {Experimental demonstration of dynamic topological
  pumping across incommensurate bilayered acoustic metamaterials},\ }\href@noop
  {} {\bibfield  {journal} {\bibinfo  {journal} {Physical Review Letters}\
  }\textbf {\bibinfo {volume} {125}},\ \bibinfo {pages} {224301} (\bibinfo
  {year} {2020})}\BibitemShut {NoStop}%
\bibitem [{\citenamefont {Spurrier}\ and\ \citenamefont
  {Cooper}(2020)}]{Spurrier2020}%
  \BibitemOpen
  \bibfield  {author} {\bibinfo {author} {\bibfnamefont {S.}~\bibnamefont
  {Spurrier}}\ and\ \bibinfo {author} {\bibfnamefont {N.~R.}\ \bibnamefont
  {Cooper}},\ }\bibfield  {title} {\bibinfo {title} {Kane-mele with a twist:
  Quasicrystalline higher-order topological insulators with fractional mass
  kinks},\ }\href {https://doi.org/10.1103/PhysRevResearch.2.033071} {\bibfield
   {journal} {\bibinfo  {journal} {Phys. Rev. Research}\ }\textbf {\bibinfo
  {volume} {2}},\ \bibinfo {pages} {033071} (\bibinfo {year}
  {2020})}\BibitemShut {NoStop}%
\bibitem [{\citenamefont {Xia}\ \emph {et~al.}(2020)\citenamefont {Xia},
  \citenamefont {Erturk},\ and\ \citenamefont {Ruzzene}}]{xia2020topological}%
  \BibitemOpen
  \bibfield  {author} {\bibinfo {author} {\bibfnamefont {Y.}~\bibnamefont
  {Xia}}, \bibinfo {author} {\bibfnamefont {A.}~\bibnamefont {Erturk}},\ and\
  \bibinfo {author} {\bibfnamefont {M.}~\bibnamefont {Ruzzene}},\ }\bibfield
  {title} {\bibinfo {title} {Topological edge states in quasiperiodic locally
  resonant metastructures},\ }\href@noop {} {\bibfield  {journal} {\bibinfo
  {journal} {Physical Review Applied}\ }\textbf {\bibinfo {volume} {13}},\
  \bibinfo {pages} {014023} (\bibinfo {year} {2020})}\BibitemShut {NoStop}%
\bibitem [{\citenamefont {Gupta}\ and\ \citenamefont
  {Ruzzene}(2020)}]{gupta2020dynamics}%
  \BibitemOpen
  \bibfield  {author} {\bibinfo {author} {\bibfnamefont {M.}~\bibnamefont
  {Gupta}}\ and\ \bibinfo {author} {\bibfnamefont {M.}~\bibnamefont
  {Ruzzene}},\ }\bibfield  {title} {\bibinfo {title} {Dynamics of quasiperiodic
  beams},\ }\href@noop {} {\bibfield  {journal} {\bibinfo  {journal}
  {Crystals}\ }\textbf {\bibinfo {volume} {10}},\ \bibinfo {pages} {1144}
  (\bibinfo {year} {2020})}\BibitemShut {NoStop}%
\bibitem [{\citenamefont {Xia}\ \emph {et~al.}(2021)\citenamefont {Xia},
  \citenamefont {Riva}, \citenamefont {Rosa}, \citenamefont {Cazzulani},
  \citenamefont {Erturk}, \citenamefont {Braghin},\ and\ \citenamefont
  {Ruzzene}}]{xia2021experimental}%
  \BibitemOpen
  \bibfield  {author} {\bibinfo {author} {\bibfnamefont {Y.}~\bibnamefont
  {Xia}}, \bibinfo {author} {\bibfnamefont {E.}~\bibnamefont {Riva}}, \bibinfo
  {author} {\bibfnamefont {M.~I.}\ \bibnamefont {Rosa}}, \bibinfo {author}
  {\bibfnamefont {G.}~\bibnamefont {Cazzulani}}, \bibinfo {author}
  {\bibfnamefont {A.}~\bibnamefont {Erturk}}, \bibinfo {author} {\bibfnamefont
  {F.}~\bibnamefont {Braghin}},\ and\ \bibinfo {author} {\bibfnamefont
  {M.}~\bibnamefont {Ruzzene}},\ }\bibfield  {title} {\bibinfo {title}
  {Experimental observation of temporal pumping in electromechanical
  waveguides},\ }\href@noop {} {\bibfield  {journal} {\bibinfo  {journal}
  {Physical Review Letters}\ }\textbf {\bibinfo {volume} {126}},\ \bibinfo
  {pages} {095501} (\bibinfo {year} {2021})}\BibitemShut {NoStop}%
\bibitem [{\citenamefont {Rosa}\ \emph
  {et~al.}(2021{\natexlab{a}})\citenamefont {Rosa}, \citenamefont {Guo},\ and\
  \citenamefont {Ruzzene}}]{rosa2021exploring}%
  \BibitemOpen
  \bibfield  {author} {\bibinfo {author} {\bibfnamefont {M.~I.}\ \bibnamefont
  {Rosa}}, \bibinfo {author} {\bibfnamefont {Y.}~\bibnamefont {Guo}},\ and\
  \bibinfo {author} {\bibfnamefont {M.}~\bibnamefont {Ruzzene}},\ }\bibfield
  {title} {\bibinfo {title} {Exploring topology of 1d quasiperiodic
  metastructures through modulated lego resonators},\ }\href@noop {} {\bibfield
   {journal} {\bibinfo  {journal} {Applied Physics Letters}\ }\textbf {\bibinfo
  {volume} {118}},\ \bibinfo {pages} {131901} (\bibinfo {year}
  {2021}{\natexlab{a}})}\BibitemShut {NoStop}%
\bibitem [{\citenamefont {Rosa}\ \emph
  {et~al.}(2021{\natexlab{b}})\citenamefont {Rosa}, \citenamefont {Ruzzene},\
  and\ \citenamefont {Prodan}}]{rosa2021topological}%
  \BibitemOpen
  \bibfield  {author} {\bibinfo {author} {\bibfnamefont {M.~I.}\ \bibnamefont
  {Rosa}}, \bibinfo {author} {\bibfnamefont {M.}~\bibnamefont {Ruzzene}},\ and\
  \bibinfo {author} {\bibfnamefont {E.}~\bibnamefont {Prodan}},\ }\bibfield
  {title} {\bibinfo {title} {Topological gaps by twisting},\ }\href@noop {}
  {\bibfield  {journal} {\bibinfo  {journal} {Communications Physics}\ }\textbf
  {\bibinfo {volume} {4}},\ \bibinfo {pages} {1} (\bibinfo {year}
  {2021}{\natexlab{b}})}\BibitemShut {NoStop}%
\bibitem [{\citenamefont {Koshino}\ and\ \citenamefont
  {Oka}(2022)}]{Koshino2022}%
  \BibitemOpen
  \bibfield  {author} {\bibinfo {author} {\bibfnamefont {M.}~\bibnamefont
  {Koshino}}\ and\ \bibinfo {author} {\bibfnamefont {H.}~\bibnamefont {Oka}},\
  }\bibfield  {title} {\bibinfo {title} {Topological invariants in
  two-dimensional quasicrystals},\ }\href
  {https://doi.org/10.1103/PhysRevResearch.4.013028} {\bibfield  {journal}
  {\bibinfo  {journal} {Phys. Rev. Research}\ }\textbf {\bibinfo {volume}
  {4}},\ \bibinfo {pages} {013028} (\bibinfo {year} {2022})}\BibitemShut
  {NoStop}%
\bibitem [{\citenamefont {Ruzzene}\ \emph {et~al.}(2003)\citenamefont
  {Ruzzene}, \citenamefont {Scarpa},\ and\ \citenamefont
  {Soranna}}]{Ruzzene2003}%
  \BibitemOpen
  \bibfield  {author} {\bibinfo {author} {\bibfnamefont {M.}~\bibnamefont
  {Ruzzene}}, \bibinfo {author} {\bibfnamefont {F.}~\bibnamefont {Scarpa}},\
  and\ \bibinfo {author} {\bibfnamefont {F.}~\bibnamefont {Soranna}},\
  }\bibfield  {title} {\bibinfo {title} {Wave beaming effects in
  two-dimensional cellular structures},\ }\href
  {https://doi.org/10.1088/0964-1726/12/3/307} {\bibfield  {journal} {\bibinfo
  {journal} {Smart Materials and Structures}\ }\textbf {\bibinfo {volume}
  {12}},\ \bibinfo {pages} {363} (\bibinfo {year} {2003})}\BibitemShut
  {NoStop}%
\bibitem [{\citenamefont {Ruzzene}\ and\ \citenamefont
  {Scarpa}(2005)}]{RuzzeneScarpa2005}%
  \BibitemOpen
  \bibfield  {author} {\bibinfo {author} {\bibfnamefont {M.}~\bibnamefont
  {Ruzzene}}\ and\ \bibinfo {author} {\bibfnamefont {F.}~\bibnamefont
  {Scarpa}},\ }\bibfield  {title} {\bibinfo {title} {Directional and band-gap
  behavior of periodic auxetic lattices},\ }\href
  {https://doi.org/https://doi.org/10.1002/pssb.200460385} {\bibfield
  {journal} {\bibinfo  {journal} {Physica status solidi (B)}\ }\textbf
  {\bibinfo {volume} {242}},\ \bibinfo {pages} {665} (\bibinfo {year}
  {2005})}\BibitemShut {NoStop}%
\bibitem [{\citenamefont {Phani}\ \emph {et~al.}(2006)\citenamefont {Phani},
  \citenamefont {Woodhouse},\ and\ \citenamefont {Fleck}}]{Phani2006}%
  \BibitemOpen
  \bibfield  {author} {\bibinfo {author} {\bibfnamefont {A.~S.}\ \bibnamefont
  {Phani}}, \bibinfo {author} {\bibfnamefont {J.}~\bibnamefont {Woodhouse}},\
  and\ \bibinfo {author} {\bibfnamefont {N.~A.}\ \bibnamefont {Fleck}},\
  }\bibfield  {title} {\bibinfo {title} {Wave propagation in two-dimensional
  periodic lattices},\ }\href {https://doi.org/10.1121/1.2179748} {\bibfield
  {journal} {\bibinfo  {journal} {The Journal of the Acoustical Society of
  America}\ }\textbf {\bibinfo {volume} {119}},\ \bibinfo {pages} {1995}
  (\bibinfo {year} {2006})}\BibitemShut {NoStop}%
\bibitem [{\citenamefont {Gonella}\ and\ \citenamefont
  {Ruzzene}(2008)}]{Gonella2008}%
  \BibitemOpen
  \bibfield  {author} {\bibinfo {author} {\bibfnamefont {S.}~\bibnamefont
  {Gonella}}\ and\ \bibinfo {author} {\bibfnamefont {M.}~\bibnamefont
  {Ruzzene}},\ }\bibfield  {title} {\bibinfo {title} {Analysis of in-plane wave
  propagation in hexagonal and re-entrant lattices},\ }\href
  {https://doi.org/https://doi.org/10.1016/j.jsv.2007.10.033} {\bibfield
  {journal} {\bibinfo  {journal} {Journal of Sound and Vibration}\ }\textbf
  {\bibinfo {volume} {312}},\ \bibinfo {pages} {125} (\bibinfo {year}
  {2008})}\BibitemShut {NoStop}%
\bibitem [{\citenamefont {Trainiti}\ \emph {et~al.}(2016)\citenamefont
  {Trainiti}, \citenamefont {Rimoli},\ and\ \citenamefont
  {Ruzzene}}]{Trainiti2016}%
  \BibitemOpen
  \bibfield  {author} {\bibinfo {author} {\bibfnamefont {G.}~\bibnamefont
  {Trainiti}}, \bibinfo {author} {\bibfnamefont {J.}~\bibnamefont {Rimoli}},\
  and\ \bibinfo {author} {\bibfnamefont {M.}~\bibnamefont {Ruzzene}},\
  }\bibfield  {title} {\bibinfo {title} {Wave propagation in undulated
  structural lattices},\ }\href
  {https://doi.org/https://doi.org/10.1016/j.ijsolstr.2016.07.006} {\bibfield
  {journal} {\bibinfo  {journal} {International Journal of Solids and
  Structures}\ }\textbf {\bibinfo {volume} {97-98}},\ \bibinfo {pages} {431}
  (\bibinfo {year} {2016})}\BibitemShut {NoStop}%
\bibitem [{\citenamefont {Beli}\ \emph {et~al.}(2018)\citenamefont {Beli},
  \citenamefont {Arruda},\ and\ \citenamefont {Ruzzene}}]{Beli2018}%
  \BibitemOpen
  \bibfield  {author} {\bibinfo {author} {\bibfnamefont {D.}~\bibnamefont
  {Beli}}, \bibinfo {author} {\bibfnamefont {J.}~\bibnamefont {Arruda}},\ and\
  \bibinfo {author} {\bibfnamefont {M.}~\bibnamefont {Ruzzene}},\ }\bibfield
  {title} {\bibinfo {title} {Wave propagation in elastic metamaterial beams and
  plates with interconnected resonators},\ }\href
  {https://doi.org/https://doi.org/10.1016/j.ijsolstr.2018.01.027} {\bibfield
  {journal} {\bibinfo  {journal} {International Journal of Solids and
  Structures}\ }\textbf {\bibinfo {volume} {139-140}},\ \bibinfo {pages} {105 }
  (\bibinfo {year} {2018})}\BibitemShut {NoStop}%
\bibitem [{\citenamefont {Foehr}\ \emph {et~al.}(2018)\citenamefont {Foehr},
  \citenamefont {Bilal}, \citenamefont {Huber},\ and\ \citenamefont
  {Daraio}}]{Foehr2018}%
  \BibitemOpen
  \bibfield  {author} {\bibinfo {author} {\bibfnamefont {A.}~\bibnamefont
  {Foehr}}, \bibinfo {author} {\bibfnamefont {O.~R.}\ \bibnamefont {Bilal}},
  \bibinfo {author} {\bibfnamefont {S.~D.}\ \bibnamefont {Huber}},\ and\
  \bibinfo {author} {\bibfnamefont {C.}~\bibnamefont {Daraio}},\ }\bibfield
  {title} {\bibinfo {title} {Spiral-based phononic plates: From wave beaming to
  topological insulators},\ }\href
  {https://doi.org/10.1103/PhysRevLett.120.205501} {\bibfield  {journal}
  {\bibinfo  {journal} {Phys. Rev. Lett.}\ }\textbf {\bibinfo {volume} {120}},\
  \bibinfo {pages} {205501} (\bibinfo {year} {2018})}\BibitemShut {NoStop}%
\bibitem [{\citenamefont {Rosi}\ and\ \citenamefont
  {Auffray}(2019)}]{Rosi2019}%
  \BibitemOpen
  \bibfield  {author} {\bibinfo {author} {\bibfnamefont {G.}~\bibnamefont
  {Rosi}}\ and\ \bibinfo {author} {\bibfnamefont {N.}~\bibnamefont {Auffray}},\
  }\bibfield  {title} {\bibinfo {title} {Continuum modelling of frequency
  dependent acoustic beam focussing and steering in hexagonal lattices},\
  }\href {https://doi.org/https://doi.org/10.1016/j.euromechsol.2019.103803}
  {\bibfield  {journal} {\bibinfo  {journal} {European Journal of Mechanics -
  A/Solids}\ }\textbf {\bibinfo {volume} {77}},\ \bibinfo {pages} {103803}
  (\bibinfo {year} {2019})}\BibitemShut {NoStop}%
\bibitem [{\citenamefont {Grabec}\ \emph {et~al.}(2020)\citenamefont {Grabec},
  \citenamefont {Koller}, \citenamefont {Sedlák}, \citenamefont {Kruisová},
  \citenamefont {Román-Manso}, \citenamefont {Belmonte}, \citenamefont
  {Miranzo},\ and\ \citenamefont {Seiner}}]{Grabec2020}%
  \BibitemOpen
  \bibfield  {author} {\bibinfo {author} {\bibfnamefont {T.}~\bibnamefont
  {Grabec}}, \bibinfo {author} {\bibfnamefont {M.}~\bibnamefont {Koller}},
  \bibinfo {author} {\bibfnamefont {P.}~\bibnamefont {Sedlák}}, \bibinfo
  {author} {\bibfnamefont {A.}~\bibnamefont {Kruisová}}, \bibinfo {author}
  {\bibfnamefont {B.}~\bibnamefont {Román-Manso}}, \bibinfo {author}
  {\bibfnamefont {M.}~\bibnamefont {Belmonte}}, \bibinfo {author}
  {\bibfnamefont {P.}~\bibnamefont {Miranzo}},\ and\ \bibinfo {author}
  {\bibfnamefont {H.}~\bibnamefont {Seiner}},\ }\bibfield  {title} {\bibinfo
  {title} {Frequency-dependent acoustic energy focusing in hexagonal ceramic
  micro-scaffolds},\ }\href
  {https://doi.org/https://doi.org/10.1016/j.wavemoti.2019.102417} {\bibfield
  {journal} {\bibinfo  {journal} {Wave Motion}\ }\textbf {\bibinfo {volume}
  {92}},\ \bibinfo {pages} {102417} (\bibinfo {year} {2020})}\BibitemShut
  {NoStop}%
\bibitem [{\citenamefont {Lubensky}(1988)}]{Lubensky1988}%
  \BibitemOpen
  \bibfield  {author} {\bibinfo {author} {\bibfnamefont {T.}~\bibnamefont
  {Lubensky}},\ }\bibfield  {title} {\bibinfo {title} {Chapter 6 - symmetry,
  elasticity, and hydrodynamics in quasiperiodic structures},\ }in\ \href
  {https://doi.org/https://doi.org/10.1016/B978-0-12-040601-2.50011-1} {\emph
  {\bibinfo {booktitle} {Introduction to Quasicrystals}}},\ \bibinfo {series}
  {Aperiodicity and Order}, Vol.~\bibinfo {volume} {1},\ \bibinfo {editor}
  {edited by\ \bibinfo {editor} {\bibfnamefont {M.~V.}\ \bibnamefont
  {Jarić}}}\ (\bibinfo  {publisher} {Elsevier},\ \bibinfo {year} {1988})\ pp.\
  \bibinfo {pages} {199 -- 280}\BibitemShut {NoStop}%
\bibitem [{\citenamefont {Widom}(2008)}]{Widom2008}%
  \BibitemOpen
  \bibfield  {author} {\bibinfo {author} {\bibfnamefont {M.}~\bibnamefont
  {Widom}},\ }\bibfield  {title} {\bibinfo {title} {Discussion of phasons in
  quasicrystals and their dynamics},\ }\href
  {https://doi.org/10.1080/14786430802247163} {\bibfield  {journal} {\bibinfo
  {journal} {Philosophical Magazine}\ }\textbf {\bibinfo {volume} {88}},\
  \bibinfo {pages} {2339} (\bibinfo {year} {2008})}\BibitemShut {NoStop}%
\bibitem [{\citenamefont {Ra'di}\ \emph {et~al.}(2017)\citenamefont {Ra'di},
  \citenamefont {Sounas},\ and\ \citenamefont {Al\`u}}]{Radi2017}%
  \BibitemOpen
  \bibfield  {author} {\bibinfo {author} {\bibfnamefont {Y.}~\bibnamefont
  {Ra'di}}, \bibinfo {author} {\bibfnamefont {D.~L.}\ \bibnamefont {Sounas}},\
  and\ \bibinfo {author} {\bibfnamefont {A.}~\bibnamefont {Al\`u}},\ }\bibfield
   {title} {\bibinfo {title} {Metagratings: Beyond the limits of graded
  metasurfaces for wave front control},\ }\href
  {https://doi.org/10.1103/PhysRevLett.119.067404} {\bibfield  {journal}
  {\bibinfo  {journal} {Phys. Rev. Lett.}\ }\textbf {\bibinfo {volume} {119}},\
  \bibinfo {pages} {067404} (\bibinfo {year} {2017})}\BibitemShut {NoStop}%
\bibitem [{\citenamefont {Popov}\ \emph {et~al.}(2018)\citenamefont {Popov},
  \citenamefont {Boust},\ and\ \citenamefont {Burokur}}]{Popov2018}%
  \BibitemOpen
  \bibfield  {author} {\bibinfo {author} {\bibfnamefont {V.}~\bibnamefont
  {Popov}}, \bibinfo {author} {\bibfnamefont {F.}~\bibnamefont {Boust}},\ and\
  \bibinfo {author} {\bibfnamefont {S.~N.}\ \bibnamefont {Burokur}},\
  }\bibfield  {title} {\bibinfo {title} {Controlling diffraction patterns with
  metagratings},\ }\href {https://doi.org/10.1103/PhysRevApplied.10.011002}
  {\bibfield  {journal} {\bibinfo  {journal} {Phys. Rev. Applied}\ }\textbf
  {\bibinfo {volume} {10}},\ \bibinfo {pages} {011002} (\bibinfo {year}
  {2018})}\BibitemShut {NoStop}%
\bibitem [{\citenamefont {Pasqual}\ \emph {et~al.}(2010)\citenamefont
  {Pasqual}, \citenamefont {Herzog},\ and\ \citenamefont
  {Arruda}}]{Pasqual2010}%
  \BibitemOpen
  \bibfield  {author} {\bibinfo {author} {\bibfnamefont {A.~M.}\ \bibnamefont
  {Pasqual}}, \bibinfo {author} {\bibfnamefont {P.}~\bibnamefont {Herzog}},\
  and\ \bibinfo {author} {\bibfnamefont {J.~R. d.~F.}\ \bibnamefont {Arruda}},\
  }\bibfield  {title} {\bibinfo {title} {Theoretical and experimental analysis
  of the electromechanical behavior of a compact spherical loudspeaker array
  for directivity control},\ }\href {https://doi.org/10.1121/1.3500689}
  {\bibfield  {journal} {\bibinfo  {journal} {The Journal of the Acoustical
  Society of America}\ }\textbf {\bibinfo {volume} {128}},\ \bibinfo {pages}
  {3478} (\bibinfo {year} {2010})}\BibitemShut {NoStop}%
\end{thebibliography}%


\begin{thebibliography}{1}%
\makeatletter
\providecommand \@ifxundefined [1]{%
 \@ifx{#1\undefined}
}%
\providecommand \@ifnum [1]{%
 \ifnum #1\expandafter \@firstoftwo
 \else \expandafter \@secondoftwo
 \fi
}%
\providecommand \@ifx [1]{%
 \ifx #1\expandafter \@firstoftwo
 \else \expandafter \@secondoftwo
 \fi
}%
\providecommand \natexlab [1]{#1}%
\providecommand \enquote  [1]{``#1''}%
\providecommand \bibnamefont  [1]{#1}%
\providecommand \bibfnamefont [1]{#1}%
\providecommand \citenamefont [1]{#1}%
\providecommand \href@noop [0]{\@secondoftwo}%
\providecommand \href [0]{\begingroup \@sanitize@url \@href}%
\providecommand \@href[1]{\@@startlink{#1}\@@href}%
\providecommand \@@href[1]{\endgroup#1\@@endlink}%
\providecommand \@sanitize@url [0]{\catcode `\\12\catcode `\$12\catcode
  `\&12\catcode `\#12\catcode `\^12\catcode `\_12\catcode `\%12\relax}%
\providecommand \@@startlink[1]{}%
\providecommand \@@endlink[0]{}%
\providecommand \url  [0]{\begingroup\@sanitize@url \@url }%
\providecommand \@url [1]{\endgroup\@href {#1}{\urlprefix }}%
\providecommand \urlprefix  [0]{URL }%
\providecommand \Eprint [0]{\href }%
\providecommand \doibase [0]{https://doi.org/}%
\providecommand \selectlanguage [0]{\@gobble}%
\providecommand \bibinfo  [0]{\@secondoftwo}%
\providecommand \bibfield  [0]{\@secondoftwo}%
\providecommand \translation [1]{[#1]}%
\providecommand \BibitemOpen [0]{}%
\providecommand \bibitemStop [0]{}%
\providecommand \bibitemNoStop [0]{.\EOS\space}%
\providecommand \EOS [0]{\spacefactor3000\relax}%
\providecommand \BibitemShut  [1]{\csname bibitem#1\endcsname}%
\let\auto@bib@innerbib\@empty
\bibitem [{\citenamefont {Hussein}\ \emph {et~al.}(2014)\citenamefont
  {Hussein}, \citenamefont {Leamy},\ and\ \citenamefont
  {Ruzzene}}]{Hussein2014}%
  \BibitemOpen
  \bibfield  {author} {\bibinfo {author} {\bibfnamefont {M.~I.}\ \bibnamefont
  {Hussein}}, \bibinfo {author} {\bibfnamefont {M.~J.}\ \bibnamefont {Leamy}},\
  and\ \bibinfo {author} {\bibfnamefont {M.}~\bibnamefont {Ruzzene}},\
  }\bibfield  {title} {\bibinfo {title} {Dynamics of phononic materials and
  structures: Historical origins, recent progress, and future outlook},\ }\href
  {https://doi.org/10.1115/1.4026911} {\bibfield  {journal} {\bibinfo
  {journal} {Applied Mechanics Reviews}\ }\textbf {\bibinfo {volume} {66}},\
  \bibinfo {pages} {040802} (\bibinfo {year} {2014})}\BibitemShut {NoStop}%
\end{thebibliography}%

\end{document}


\preprint{Physical Review Materials}

\title{Supplemental Material \\ Wave beaming and diffraction in quasicrystalline elastic metamaterial plates}

\author{Danilo Beli} \email{dbeli@usp.br, beli.danilo@gmail.com}
\affiliation{Department of Aeronautical Engineering, Sao Carlos School of Engineering, University of Sao Paulo, Brazil}
\author{Matheus Inguaggiato Nora Rosa} \email{matheus.rosa@colorado.edu}
\affiliation{Department of Mechanical Engineering, College of Engineering and Applied Science, University of Colorado Boulder, USA \vspace{0.25cm}} 
\author{Carlos De Marqui Jr.} \email{demarqui@sc.usp.br}
\affiliation{Department of Aeronautical Engineering, Sao Carlos School of Engineering, University of Sao Paulo, Brazil}
\author{Massimo Ruzzene} \email{massimo.ruzzene@colorado.edu}
\affiliation{Department of Mechanical Engineering, College of Engineering and Applied Science, University of Colorado Boulder, USA \vspace{0.25cm}}


\maketitle



\section{Designs for the 4-, 6- and 8-fold symmetries} \label{sec:AppA}

\begin{figure*}
	\centering
	{\includegraphics[scale=0.490]{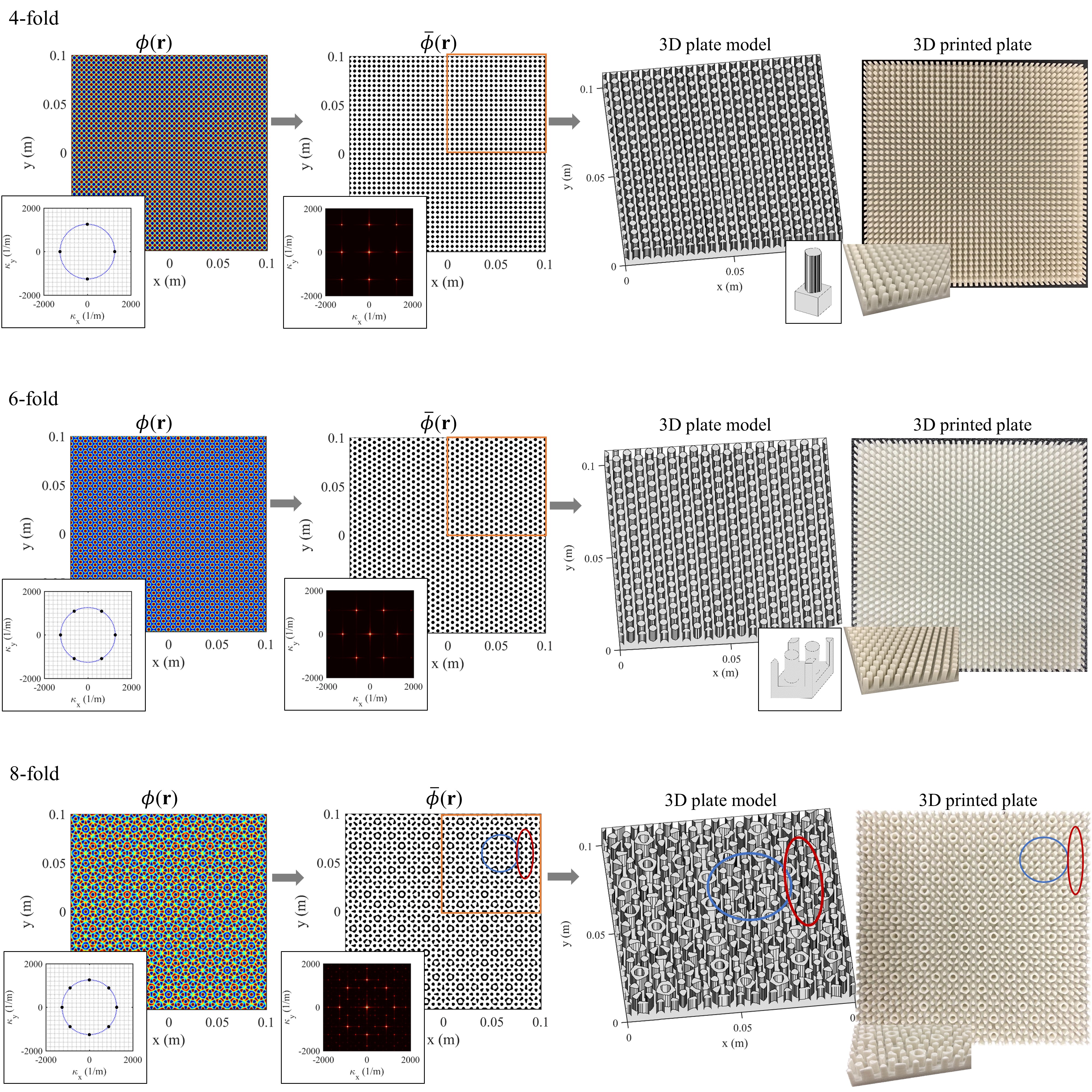}} 
	\caption{\label{fig:figureSM1}  Design strategy for the 4- (top), 6- (middle) and 8-fold (bottom) elastic metamaterial plate with vf = 0.30: two-dimensional physical distribution $\phi$ by assigning 4, 6 or 8 Bragg peaks, two-phase distribution $\bar{\phi}$ after applying the threshold procedure and its Fourier transform: phase A (white) and phase B (black). The three-dimensional plates are constructed by extruding phase B towards z direction, which produces the modulated thickness (pillars) and the correspondent 3D printed plates.} 
\end{figure*}

The design strategy to conceive the 4-, 6- and 8-fold elastic metamaterial plates by assigning the respective number of Bragg peaks in reciprocal space is illustrated in Fig. \ref{fig:figureSM1}. The geometric ($L$, $\lambda_{0}$, $h_{\tt A}$ and $h_{\tt B}$) and material ($\rho_n$, $E_n$, $\nu_n$ and $\eta_n$) properties as well as the volume fraction (${\tt vf} = 0.30$) are the same used in the 10-fold quasicrystalline metamaterial plate presented in Section II of the paper. For the periodic cases, the 3D square and hexagonal unit cells are also depicted with the respective 3D elastic plate models.


\section{Approximate Dispersion Properties} \label{sec:AppC}

\begin{figure*}
	\centering
	{\includegraphics[scale=0.400]{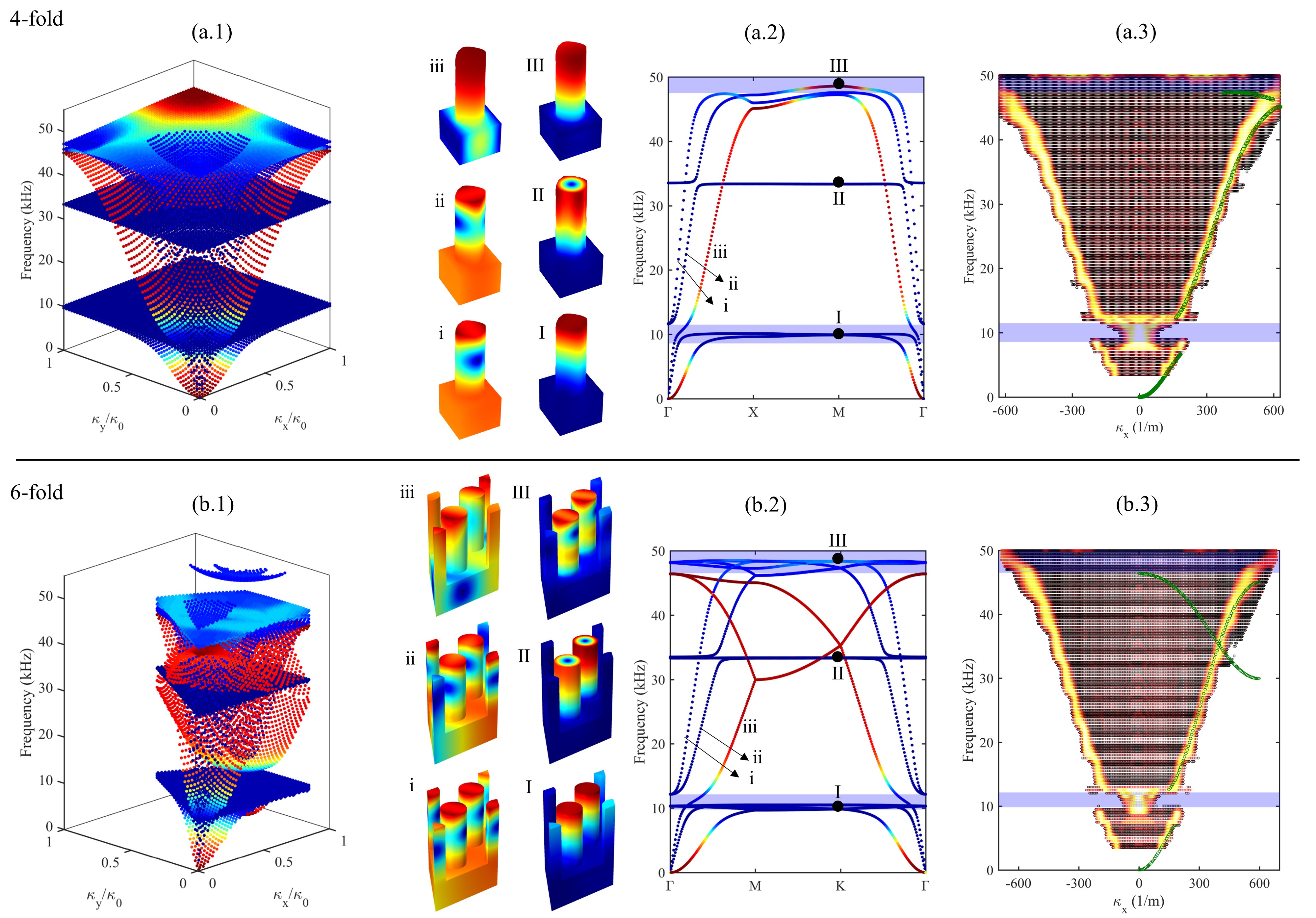}} 
	\caption{\label{fig:figureSM5}  Dispersion properties and contours at specific frequencies of the 4- (a), 6- (b) and 8-fold (c) elastic plate obtained by numerical simulations and experiments. Numerical and experimental results (3D spectrum and wave number contours) are in good agreement. For the 8-fold quasicrystalline plate, the contours at specific frequencies show the transition between two bands twisted by $\theta_N/2$.} 
\end{figure*}

The numerical and experimental approximate dispersion surfaces for the 4-, 6- and 8-fold metamaterial plates shown in Fig. \ref{fig:figureSM5} are obtained from the same procedure described in the main text: time response, Fourier transform and a wave filtering.

For both periodic plates (4-fold: a.1-a.3, and 6-fold: b.1-b.3), the numerical approximate dispersion surfaces present circular wave number contour for almost every frequency in the spectrum, revealing an isotropic dispersion. Discontinuities appear in frequency zones around 10 kHz and 50 kHz that corresponds to band gaps in the bending waves (i.e., out-of-plane motion). The approximated dispersion surfaces are also reconstructed using experimental data (the experimental set-up is described in the main text and illustrated in Fig. 2), and the results are also shown in Fig. \ref{fig:figureSM5} (4-fold: a.4-a.6, and 6-fold: b.4-b.6). The experimental dispersion surfaces also have circular wave number contours with well defined band gaps at similar frequencies. A good agreement has been achieved between numerical and experimental results in both frequency and wave number domains.  

The approximate dispersion results for the 8-fold elastic metamaterial plate are shown in Fig. \ref{fig:figureSM5} (numerical: c.1-c.3, and experimental: c.4-c.6). The propagating bands displayed in the 8-fold symmetry change the dispersion with the frequency, i.e. they twist similarly to the 10-fold case. However, only two twists are observed at high frequencies while at the low frequencies a more isotropic dispersion is observed. One of the transitions is also highlighted in the wave number contours of Fig. \ref{fig:figureSM5} (o,r), where numerical and experimental results are in a good agreement despite the frequency shift of 2 kHz. In addition, the zones with complete band gaps appear around the band gap zones of the periodic cases: 10 kHz and 50 kHz. 


\section{Dispersion Properties using Bloch Periodic Conditions} \label{sec:AppB}

For the periodic cases, the dispersion surfaces and band structures can be computed by applying Bloch periodic conditions at the unit cell boundaries which leads to the eigenvalue problem $\omega({\bf \kappa})$: 
\begin{equation}
\tag{S1}
[\hat{\bf K}(\kappa)-\omega^2(\kappa) \hat{\bf M}(\kappa)] {\bf u (\kappa)} = 0 ,
\label{eq:eqSM0}
\end{equation}
where $\kappa$ is the wave vector, $\hat{\bf K}(\kappa)$ and $\hat{\bf M}(\kappa)$ are the stiffness and mass finite element matrices of the unit cell after applying the Bloch periodic conditions \cite{Hussein2014}. For the band structure computation, $\kappa$ is swept over the correspondent first Brillouin zone contour. The bending waves, which have mainly out-of-plane motion, are distinguished by the wave polarization computation given by
\begin{equation}
\tag{S2}
p_z = \frac{\int u_z^2 dV}{\int \left( u_x^2+u_y^2+u_z^2 \right) dV} ,
\label{eq:eqSM1}
\end{equation}
where $u_i$ is the displacement of the wave mode shape at the direction $i \in [x, y, z]$. The polarization $p_z$ is linked to the bending wave and ranges from 0 (pure in-plane motion) to 1 (pure out-of-plane motion). This exact dispersion computation of the periodic plates are used to validate the approximate dispersion results obtained from the time response and its Fourier transform (3D-FT) as well as to guide the dispersion analysis in the quasicrystalline plates.

\begin{figure*}
	\centering
	{\includegraphics[scale=0.500]{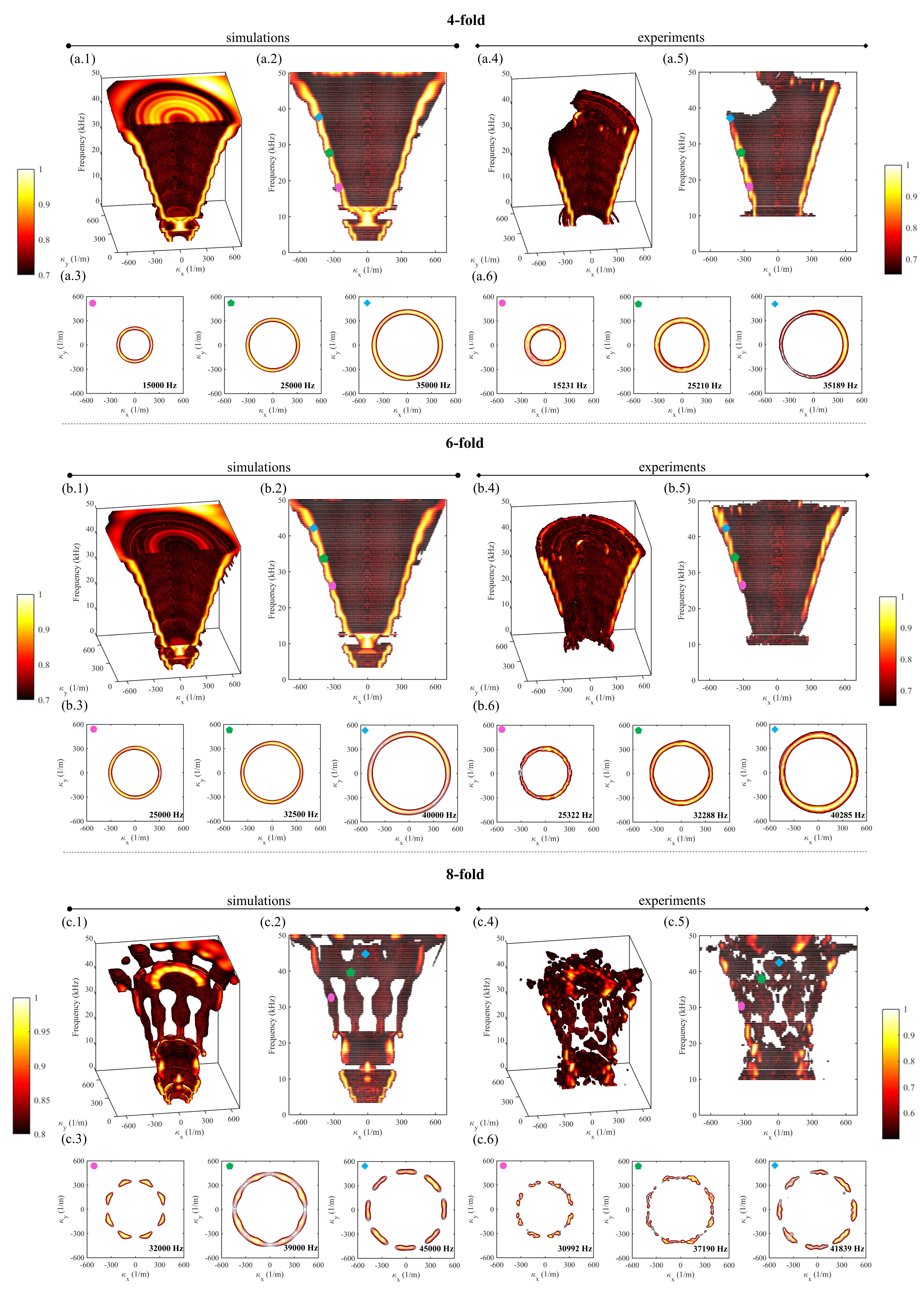}} 
	\caption{\label{fig:figureSM2} Dispersion properties for the 4-fold (a) and 6-fold (b) elastic metamaterial plates. Dispersion surface (a.1, b.1) and band structure (a.2, b.2) computed by applying Bloch-Floquet conditions at 4-fold and 6-fold unit cell boundaries, the colors represent the wave polarization: in-plane (blue) and out-of-plane (red). The wave mode shapes at each locally resonant band gap are also presented, their colors represent the displacement amplitude: low (blue) and high (red). The approximate band structure obtained by the time response and its Fourier transform (a.3, b.3) are compared to the dispersion computed by Bloch-Floquet theory (green) at $\Gamma$X and $\Gamma$K directions, respectively.} 
\end{figure*}

The computed dispersion results by Bloch-Floquet theory for the 4-fold and 6-fold plates are presented in Fig. \ref{fig:figureSM2}. Three main waves are distinguished: longitudinal (i), shear (ii) and flexural (iii) as pointed out by the wave mode shapes. Three band gaps related to the pillars resonance are opened: the first around 10 kHz is related to the flexural motion of the pillars acting in all the waves (I), the second around 31 kHz is related to torsional motion acting only on the shear wave (II) and the third around 50 kHz is related to the second flexural mode that opens a full band gap (III). The approximate dispersion surfaces computed by the time response and the 3D-FT are compared to the exact dispersion results obtained by enforcing Bloch conditions as shown in Fig. \ref{fig:figureSM2}(a.3, b.3), where a good agreement is observed in the propagating bands and band gaps for the bending wave.


\section{Computation of the Group Velocity Contours} \label{sec:AppB2}

The elastic energy flow throughout the metastructure is related to the group velocity analysis ${\bf c} = \nabla_{\kappa} \omega$, which implies that the waves propagate in directions normal to the wave number contour at a constant frequency. For this purpose, the wave number contour is estimated using the approximate dispersion results (3D-FT), and for a direction given by the angle $\alpha \in [0 \quad 2 \pi]$, the approximate wave number at the frequency $\omega_j$ is given by the weighted average: 
\begin{equation}
\tag{S3}
\kappa_{\alpha}(\omega_j) = \frac{\sum_{i} \kappa_{ri} \left[ {\bf \hat{U}}_{t}(\kappa_{ri},\omega_j) \right]^{\beta}} {\sum_{i} \left[ {\bf \hat{U}}_{t}(\kappa_{ri},\omega_j) \right]^{\beta}},
\label{eq:eqSM2}
\end{equation}
where $\beta$ is the weighted parameter applied on the normalized Fourier coefficients ${\bf \hat{U}}_{t}$ ($\beta = 5$ has been used in this work) and $\kappa_{ri} = (\kappa_{xi}^2+\kappa_{yi}^2)^{1/2}$ is the radial wave number at the direction $\alpha$. Therefore, the group velocity for an angular direction $\alpha$ is numerically computed through the finite difference:
\begin{equation}
\tag{S4}
c_{\alpha}(\omega_j) = \frac{\partial \omega_j}{\partial \kappa_{\alpha}} \approx \frac{\omega_{j+1}-\omega_j}{\kappa_{\alpha}(\omega_{j+1})-\kappa_{\alpha}(\omega_{j})},
\label{eq:eqSM3}
\end{equation}
with $\omega_{j+1} \rightarrow \omega_j$. 

\begin{figure*}
	\centering
	{\includegraphics[scale=0.50]{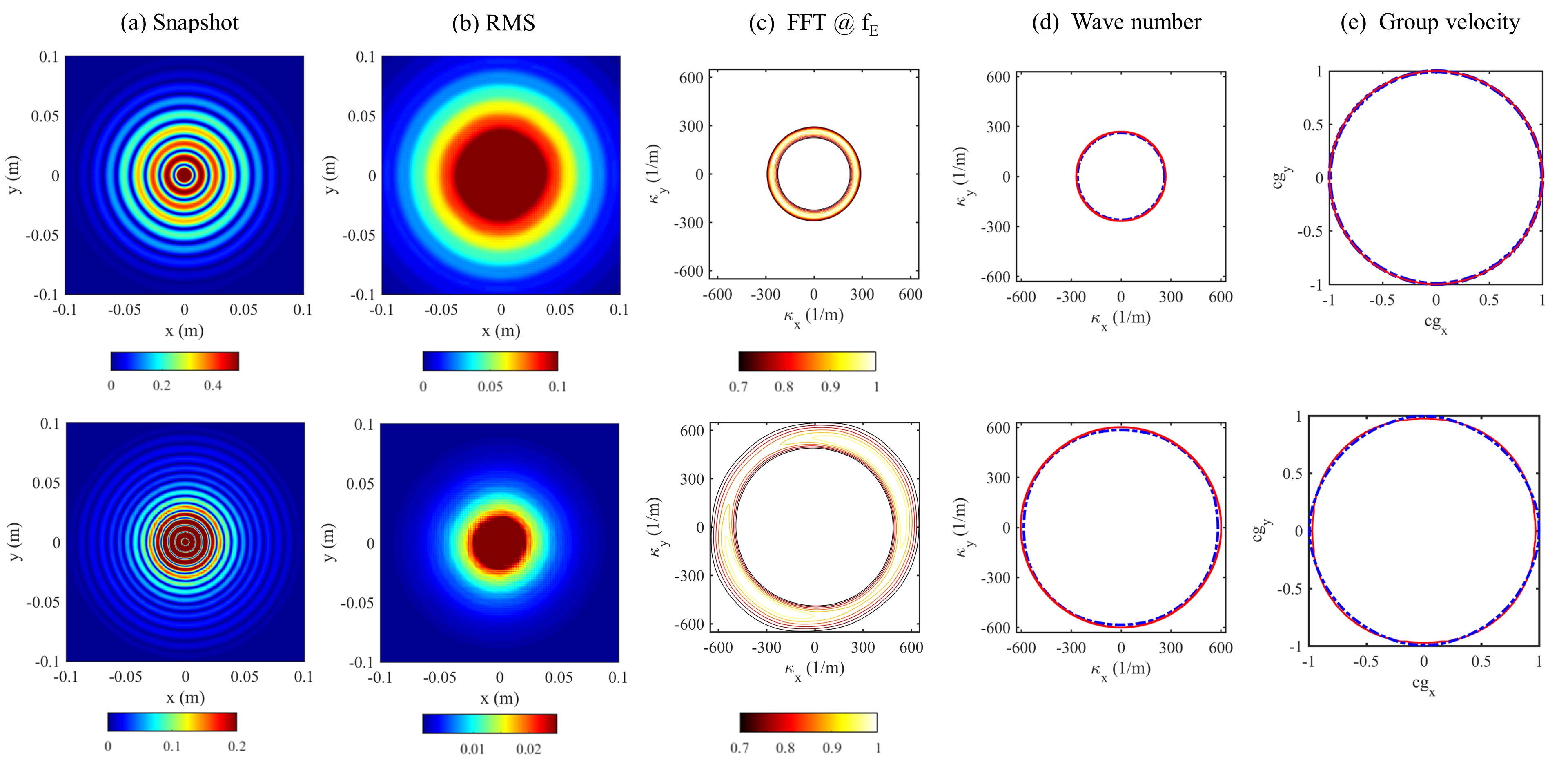}} 
	\caption{\label{fig:figureSM3} Directional wave behavior for the 4-fold metamaterial plate at 20 kHz (top) and 45 kHz (bottom). Snapshot (a) and RMS of the displacement field over time (b), the wave number contour at the excitation frequency obtained by the 3D-FT (c). The wave number contour (d) and the correspondent normalized group velocity pattern (e): approximate results obtained by the approach presented in Section II of this SM (blue dashed line) and the results computed by using the Bloch-Floquet theory (red line). A isotropic wave behavior is observed in both frequencies: 20 and 45 kHz.} 
\end{figure*}

\begin{figure*}
	\centering
	{\includegraphics[scale=0.50]{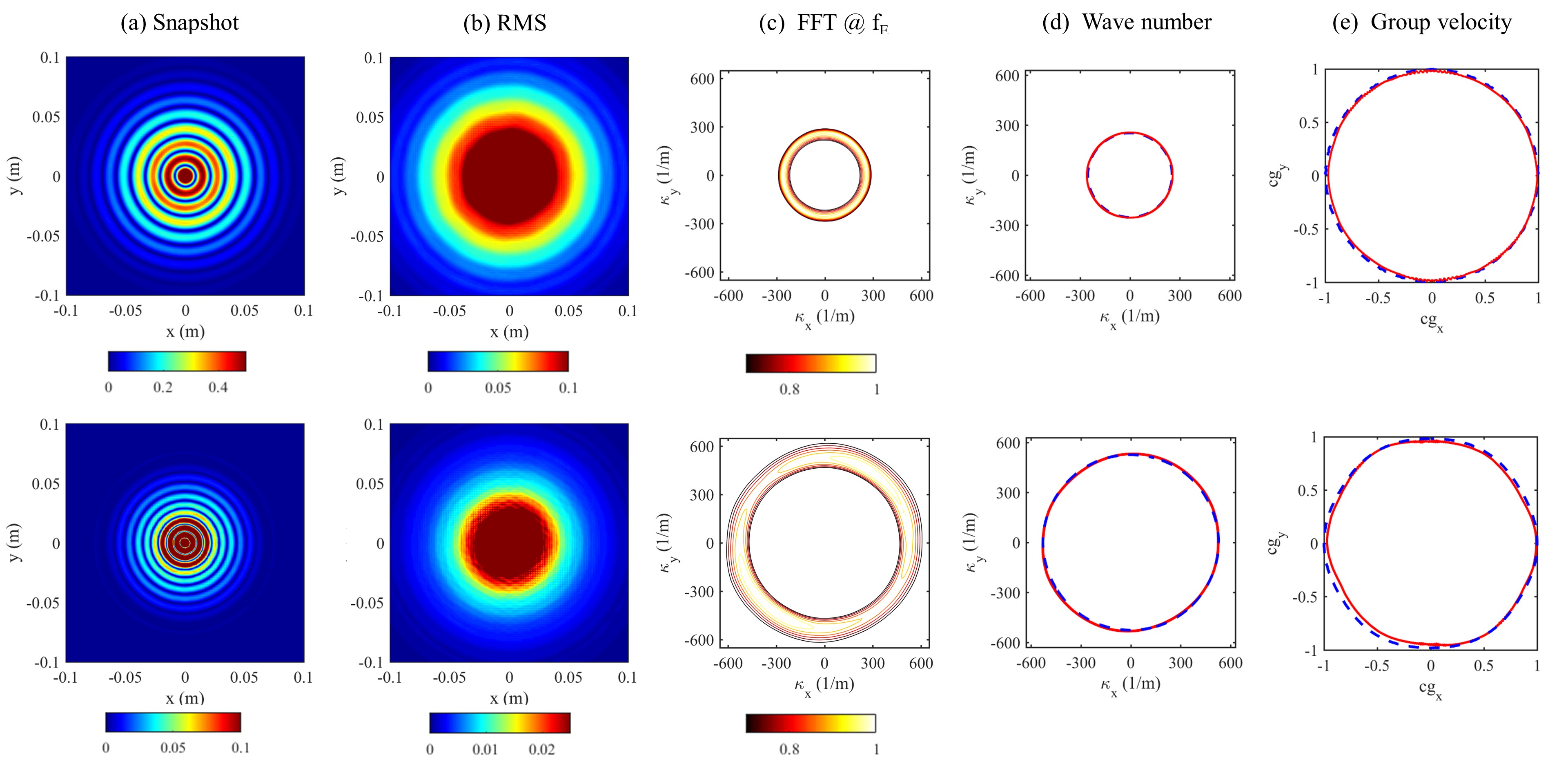}} 
	\caption{\label{fig:figureSM4} Directional wave behavior for the 6-fold metamaterial plate at 20 kHz (top) and 45 kHz (bottom). Snapshot (a) and RMS of the displacement field over time (b), the wave number contour at the excitation frequency obtained by the 3D-FT (c). The wave number contour (d) and the correspondent normalized group velocity pattern (e) : approximate results obtained by the approach presented in Section II of this SM (blue dashed line) and the results computed by using the Bloch-Floquet theory (red line). While a isotropic wave behavior is observed at 20 kHz, a low anisotropic wave behavior following the 6-fold pattern is observed at 45 kHz.} 
\end{figure*}

Figures \ref{fig:figureSM3} and \ref{fig:figureSM4} illustrate the procedure to obtain the group velocity contour for the 4-fold and 6-fold configurations, respectively. The approximate wave number contours as well as the group velocity contours obtained from the approximate dispersion surfaces (blue dashed line) are compared to the analytical dispersion surfaces computed using the Bloch-Floquet theory (red line). A good agreement between the results are achieved for the selected frequencies. In addition, the group velocity contours are correlated to the rotational symmetry and to the wave propagation patterns depicted in the snapshot and time root mean square results. While the 4-fold plate has an omnidirectional wave propagation at 20 kHz and 45 kHz, the 6-fold plate presents an isotropic and anisotropic wave propagation at 20 kHz and 45 kHz, respectively. Therefore, the approximate dispersion results computed from the time response and its Fourier transform also provide precise and rich wave information regarding the attenuation zones and propagating bands as well as allows qualitative wave directionality analyses.


\section{Wave Directionality of the 8-fold metamaterial plate} \label{sec:AppD}

The wave directionality results for the 8-fold plate are presented in this section to support the discussions and achievements regarding the anisotropic wave phenomena in quasicrystalline plates of the main text. The same procedure of the main text is employed here, i.e. a time response analysis by applying a punctual load at the plate center with burst sine shape with several cycles and free-free boundary conditions.

\begin{figure*}
	\centering
	{\includegraphics[scale=0.48]{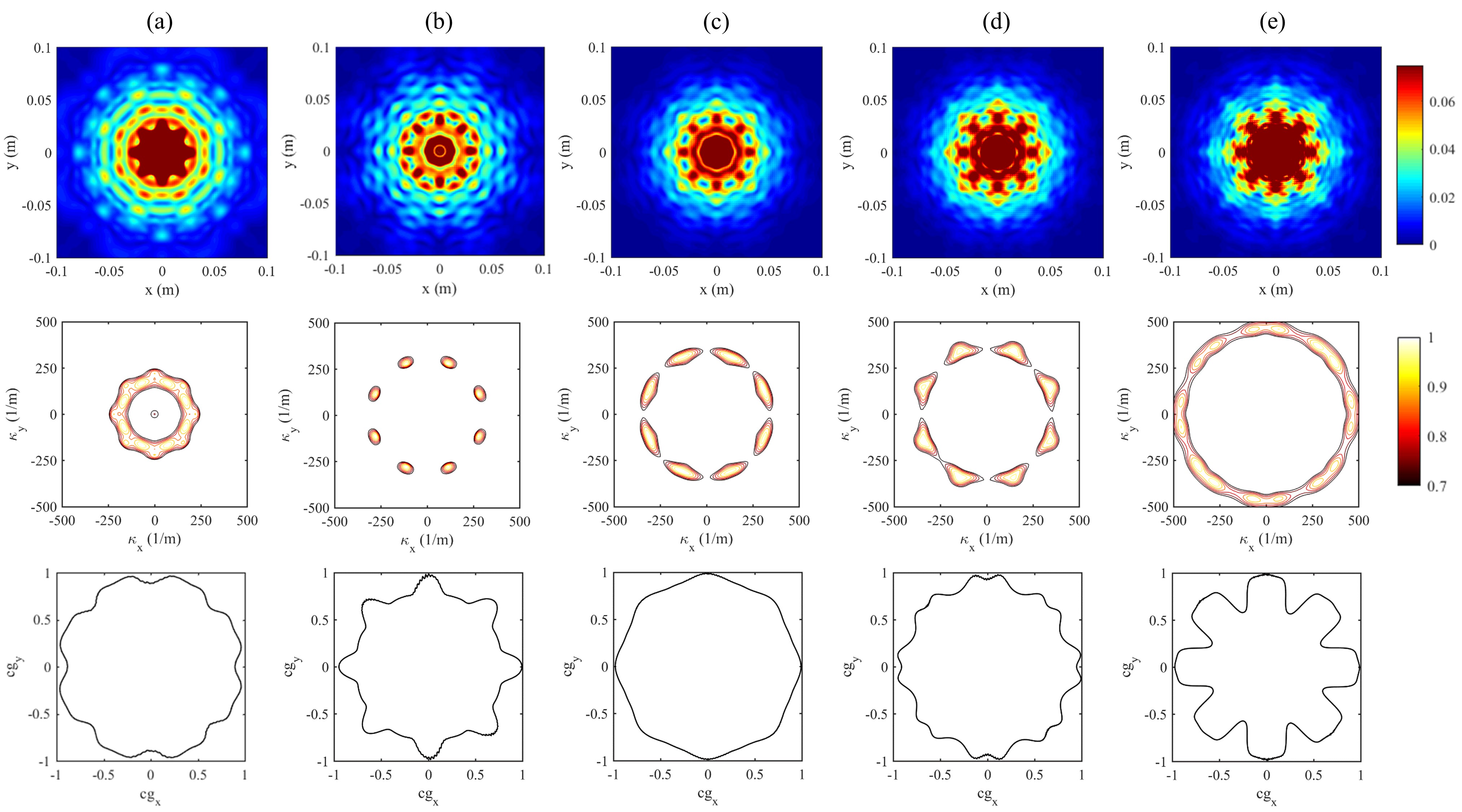}} 
	\caption{\label{fig:figureSM6} Directional wave behavior for the 8-fold metamaterial plate at different excitation frequencies: 15 kHz (a), 25 kHz (b), 30 kHz (c), 35 kHz (d) and 45 kHz (e). The first row corresponds to the RMS of the displacement field over time, the second line corresponds to the RMS of the wave number contour over frequency, and the third line corresponds to the approximate group velocity contour. The wave directionality is correlated to the rotational symmetry of the metamaterial plate.} 
\end{figure*}

The wave directionality results are shown in Fig. \ref{fig:figureSM6}. From  the center of the plate, the wave propagates branching in a 8-fold pattern correlated to its rotational symmetry. However, a frequency-dependent directionality is observed since the wave directionality twists from low to high frequencies, as confirmed by the group velocity contours. Along most of the frequency range, nearly isotropic wave propagation occurs, even at the zones with sharp Fourier peaks. Only at higher frequencies (i.e., $45$ kHz) a clear wave directionality is observed, which is independent of the almost isotropic wave number contour at the same frequency.


\section{Extra Wave Diffraction Results for the 10-fold Metamaterial Plate} \label{sec:AppE}

\begin{figure*}
	\centering
	{\includegraphics[scale=0.48]{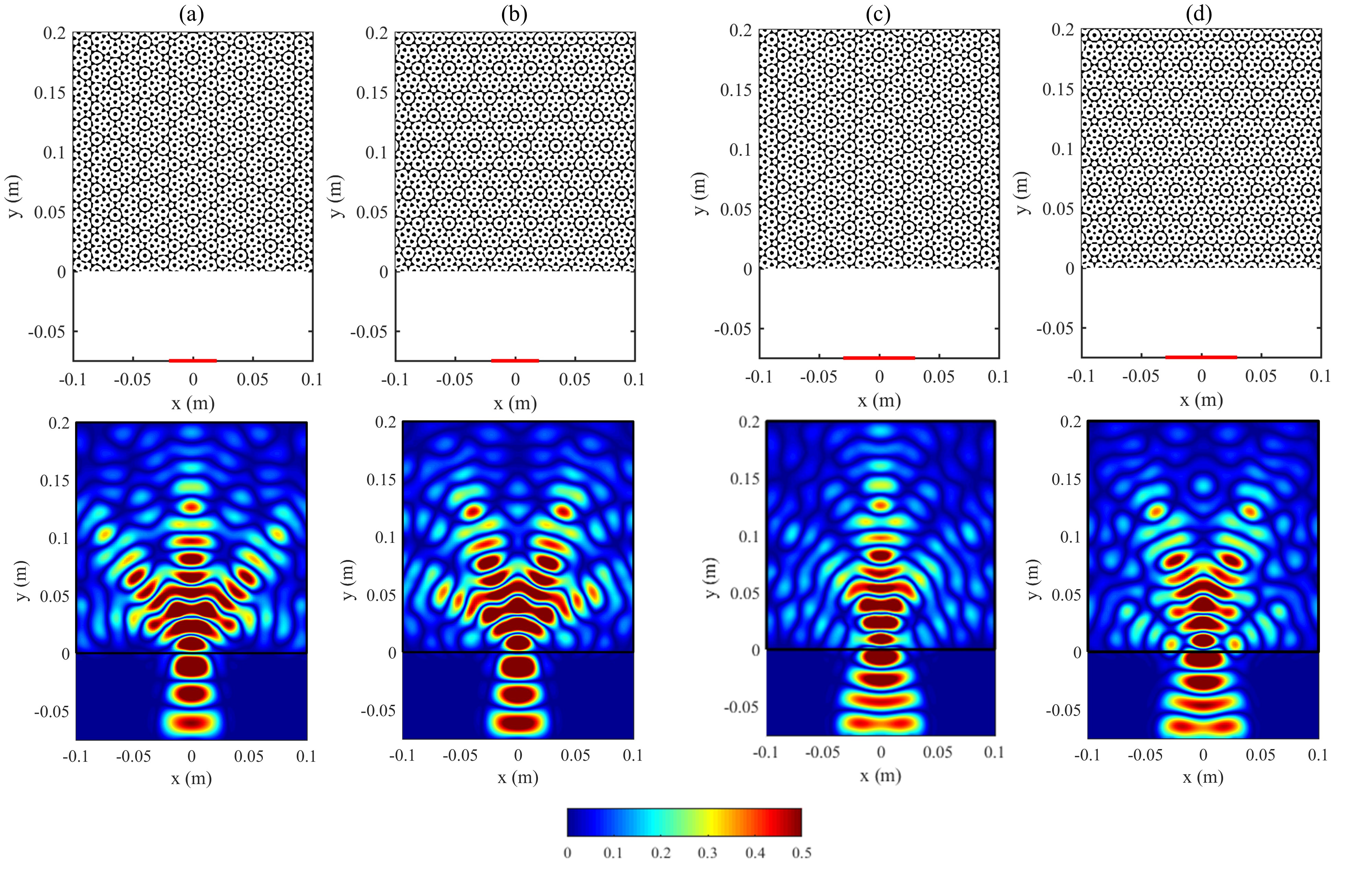}} 
	\caption{\label{fig:figureSM7} Extra diffraction results for the 10-fold metamaterial plate: extended designs with the location of excitation in red (top row) and displacement field for a snapshot after the wave front branching (bottom row). The cases (a-d) follow the description of the main text and correspondent to the cases (a-d) in Fig. 7.} 
\end{figure*}

The plate configurations and the time snapshots of the diffraction response for the 10-fold plate discussed in Section V of the main text are shown in Fig. \ref{fig:figureSM7}. A clear branching of the wave fronts in the expected directions is also observed in the snapshots.


\bibliography{references}%